\documentclass[twocolumn]{aastex702}
\usepackage{amsmath}
\usepackage{enumerate}
\usepackage{enumitem}   
\usepackage{hyperref}
\usepackage{color, colortbl}
\usepackage{subfigure}
\usepackage{color}
\usepackage{courier}
\usepackage{mathtools}
\usepackage{threeparttable}
\usepackage{enumitem}
\usepackage{rotating}
\usepackage{tabularx}
\usepackage{rotating}
\usepackage{CJK}
\usepackage{multirow}
\definecolor{H2color}{HTML}{E9D7C3}
\definecolor{Zcolor}{HTML}{BADDFF}
\definecolor{lightgrayrow}{gray}{0.92}

\newcommand{\Mstar}{$M_{\star}$} % stellar mass
\newcommand{\Reff}{$R_\mathrm{e}$} % effective radius
\newcommand{\OH}{12 $+$ $\log$(O/H)} % 12 $+$ $\log$(O/H)
\newcommand{\Dsf}{$\Delta\log(\Sigma_{\rm SFR})$}
\newcommand{\Doh}{$\Delta[12+\log(\mathrm{O/H})]$}

\begin{document}

\title{Dependence of Merger-driven Star Formation and Gas-phase Metallicity Evolution on Merger Type and Stellar Mass}

\begin{CJK}{UTF8}{bkai}

\author[orcid=0000-0002-1370-6964,sname=Pan]{Hsi-An Pan (潘璽安)}
\affiliation{Department of Physics, Tamkang University, No.151, Yingzhuan Road, Tamsui District, New Taipei City 251301, Taiwan}
\email[show]{hapan@gms.tku.edu.tw}

\begin{abstract}
Galaxy mergers play a fundamental role in driving gas inflows, altering star formation, and redistributing metals, yet how these processes depend on merger type, merger stage, and progenitor galaxy properties remains incompletely understood. In this work, we investigate the resolved impact of mergers on star formation rate (SFR) and gas-phase metallicity using MaNGA integral-field spectroscopy and the MaNGA Galaxy Merger Catalog of \citet{Com24}. By comparing merger galaxies to matched controls, we quantify local offsets in SFR and metallicity as functions of merger type, merger stage, stellar mass, and galactocentric radius.
We find that major mergers generally exhibit stronger SFR enhancement and greater metallicity dilution than minor mergers.  In major mergers, SFR enhancement is often spatially extended, whereas in minor mergers it is weaker and more centrally concentrated. The SFR response in major mergers varies non-monotonically with merger stage, with stronger enhancement in the early (first-passage) and post-coalescence stages and weaker enhancement in the late (between-passage) stage. Metallicity dilution is generally strongest in early-stage major mergers and weaker at later stages, although this stage dependence is not monotonic across all stellar-mass bins. We also find stellar-mass dependence that lower-mass systems show stronger metallicity dilution in both major and minor mergers, while the strongest SFR enhancement occurs in intermediate- and high-mass major mergers. These results show that the resolved star formation and metallicity responses depend jointly on merger type, merger stage, and stellar mass and exhibit different behaviors across the merger sequence.

\end{abstract}



\section{Introduction}

Galaxy mergers are a fundamental process in galaxy evolution, capable of reshaping galaxy structure, redistributing angular momentum, and altering the interstellar medium (ISM) (e.g., \citealt{Mih96,Hop08}). Theoretical and numerical studies predict that tidal torques generated during mergers can remove angular momentum from the gas, driving large-scale inflows toward the central regions of galaxies \citep[e.g.,][]{Mih96,Ion04,Tor12,Bus18,Mor19,Mor21}. These inflows can fuel bursts of star formation \citep{Tor12,Bus18,Per25}, feed active galactic nuclei (AGN) \citep{Hop09,Cap15}, and redistribute metals throughout the disk \citep{Rup10,Tis22,Ren25,Tap25,Tis25}, making galaxy mergers one of the key mechanisms regulating the baryon cycle and driving galaxy transformation over cosmic time.

Observationally, numerous studies have shown that galaxy mergers are associated with enhanced star formation activity compared to isolated galaxies \citep[e.g.,][]{Scu12,Pat13,Per19,Hor21,Cal24}. On resolved scales, this enhancement can extend beyond the nuclear region and into the disk, presumably due the large-scale redistribution of gas during interactions \citep[e.g.,][]{Cor17,Pan19,Tho19,Tho22,Shi26}. At the same time, mergers are also linked to gas-phase metallicity dilution, where inflows of metal-poor gas from the outer disk reduce the central metallicity \citep[e.g.,][]{Scu12,Pan25,Ren25,Shi26}. These observations support the picture that star formation enhancement and metallicity dilution are physically coupled consequences of merger-driven gas inflows.

A growing number of studies have shown that these merger-driven effects evolve across the merger sequence. Both observations and simulations indicate that star formation enhancement is not constant, but instead varies with merger stage, often peaking near close passages and final coalescence \citep{Tor12,Scu12,Bus18,Pan19}. Similarly, metallicity dilution is thought to be strongest shortly after inflow episodes and to weaken as merger-triggered star formation and chemical enrichment proceed \citep{Tor19,Tis22,Pan25,Tis25,Tap25,Shi26}. These results suggest that the timing of gas inflows, star formation, and chemical enrichment are closely linked, but may not evolve synchronously (Pan et al. submitted).

However, beyond merger stage, the strength of merger-driven perturbations is also expected to depend on the mass ratio of the interacting galaxies. Simulations predict that major mergers (i.e., mergers between galaxies of comparable stellar mass) generate stronger tidal torques and more efficient gas inflows than minor mergers (i.e., mergers with unequal stellar masses), leading to stronger starbursts and larger chemical perturbations \citep[e.g.,][]{Cox08,Bus18}. In addition, galaxy stellar mass may regulate the response to mergers through its connection to gas fraction, metallicity gradients, and gravitational potential depth. 
Despite these theoretical expectations, observational constraints on how merger-driven star formation and metallicity jointly depend on merger type, merger stage, and stellar mass remain limited. 
Identifying galaxy mergers and characterizing their merger configurations in large surveys requires high-quality imaging with sufficient spatial resolution and depth to detect tidal features, while traditional visual classifications are inherently subjective and difficult to reproduce consistently across large samples \citep[see the discussion by][]{Mar24}. 
These challenges have motivated the development of automated, machine-learning-based merger classification methods.

In this work, we investigate the resolved impact of galaxy mergers on star formation and gas-phase metallicity using the Mapping Nearby Galaxies at Apache Point Observatory \citep[MaNGA;][]{Bun15} survey and the MaNGA Galaxy Merger Catalog \citep{Nev23,Com24}. The merger catalog is constructed using a supervised machine-learning framework. This enables a uniform and physically motivated classification of merger type and merger stage across a large galaxy sample. 
By constructing matched control samples at both the galaxy and spaxel levels, we quantify local offsets in star formation and metallicity relative to non-merger analogs. Unlike previous studies that primarily focused on merger stage, our analysis jointly examines the dependence on merger stage, merger type (major versus minor), and stellar mass. The goal is to disentangle how these factors regulate merger-driven gas inflows and shape their impact on the ISM.

This paper is organized as follows. Section~\ref{sec_data} describes the construction of the galaxy sample, including the derivation of star formation rates and gas-phase metallicities from the MaNGA survey, as well as the identification and classification of mergers by type and stage. Section~\ref{sec_result} presents the resolved changes in star formation and metallicity as functions of merger stage, merger type, and progenitor stellar mass. Section~\ref{sec_discuss} discusses the implications and caveats of our results, and Section~\ref{sec_summary} summarizes our main conclusions. Throughout this paper, we adopt a cosmology with $H_0 = 70$ km s$^{-1}$ Mpc$^{-1}$, $\Omega_{\rm m}=0.3$, and $\Omega_{\Lambda}=0.7$.

\section{Merger Sample Construction and Control Matching}
\label{sec_data}
\subsection{MaNGA Data and Resolved Physical Measurements}

In this work, we use data from the SDSS-IV MaNGA survey (Mapping Nearby Galaxies at Apache Point Observatory; \citealt{Bun15}) to investigate the resolved impact of galaxy mergers on star formation and  metallicity, traced by the gas-phase oxygen abundance, \OH. 
MaNGA provides integral-field spectroscopy for nearby galaxies, enabling spatially resolved measurements of emission lines and stellar continuum across each system. The survey covers a wavelength range of 3600-10300~\AA\ with a spectral resolution of $R\sim1400$ at 4000~\AA\ and $R\sim2600$ at 9000~\AA. The typical point-spread function is $\sim2\farcs5$, corresponding to $\sim1.8$ kpc at the median redshift of the sample ($z\sim0.03$).

We use the full MaNGA sample and adopt the \texttt{Pipe3D} value-added data products \citep{San16a,San16b,San18} to derive both spatially resolved and global galaxy properties. \texttt{Pipe3D} performs stellar continuum fitting and nebular emission-line measurements for each spaxel, from which we derive local star formation rates (SFRs) and \OH. 

The \texttt{Pipe3D} pipeline derives stellar population properties by fitting the observed spectra with a library of simple stellar population  models. To improve the reliability of the continuum fitting, neighboring spaxels are first adaptively binned to achieve a target continuum signal-to-noise ratio of 50. The SSP fitting is then performed on the coadded spectra within each spatial bin. The resulting best-fit stellar population model is subsequently rescaled to the continuum level of individual spaxels with continuum S/N $>3$, allowing stellar population properties to be recovered at the native spatial resolution of the data. Stellar masses are derived from these resolved stellar population models and converted into stellar mass surface densities, $\Sigma_{\star}$, by normalizing to the physical area of each spaxel, expressed in units of M$_{\odot}$~kpc$^{-2}$.

Resolved SFRs are derived from extinction-corrected H$\alpha$ emission. The extinction-corrected H$\alpha$ luminosity is converted into SFR surface density ($\Sigma_{\rm SFR}$; M$_{\odot}$~yr$^{-1}$~kpc$^{-2}$) using the calibration of \citet{Ken98}. Since reliable SFR and metallicity measurements require ionization dominated by star formation, we restrict our analysis to star-forming spaxels selected using standard Baldwin-Phillips-Terlevich (BPT) \citep{Bal81} diagnostic diagrams \citep{Kew01,Kau03,Cid13} and an H$\alpha$ equivalent width threshold of $>6$~\AA\ \citep{San14}. In addition, we require a signal-to-noise ratio greater than 3 for all emission lines used in the SFR and metallicity measurements. Gas-phase metallicities are determined using the \emph{O3N2} calibration\footnote{The \emph{O3N2} index is defined as $\mathrm{O3N2} = ([\mathrm{O,III}]\lambda5007/\mathrm{H}\beta)/([\mathrm{N,II}]\lambda6584/\mathrm{H}\alpha)$.} of \citet{Mar13}. The global SFR and \Mstar\ are derived by integrating the resolved measurements \citep{San18}. These meareuemtns are identical to those adopted in our previous studies of star formation and metallicity in MaNGA mergers \citep{Pan19,Pan25}.
The MaNGA survey is designed to map galaxies out to at least 1.5~\Reff. To ensure a fare comparison across the sample, we restrict our analysis to the common radial coverage within 1.5~\Reff.

\subsection{Merger Classification}
\label{sec_merger_classification}
To identify galaxy mergers, we use the MaNGA Galaxy Merger Catalog of \citet{Com24}, which applies the supervised machine-learning merger classification framework developed by \citet{Nev23} to MaNGA galaxies. This framework uses a linear discriminant analysis classifier based on seven morphological predictors measured from galaxy images, including the Gini coefficient, $M_{20}$, concentration, asymmetry, clumpiness, S\'{e}rsic index, and shape asymmetry. The classifier was trained on mock SDSS $r$-band images generated from hydrodynamical simulations of merging and isolated galaxies \citep{Nev19}, combined with dust radiative-transfer modeling, and was designed to distinguish mergers from non-mergers as well as to classify merger systems by mass ratio and evolutionary stage. Each merger class is assigned an independent probability, allowing galaxies to be probabilistically classified across different merger types and stages. 
The merger classification is assigned to each MaNGA-targeted galaxy. Accordingly, all resolved measurements analyzed in this work are those of the MaNGA-targeted galaxy. For most merger systems, the companion galaxy is not included in the MaNGA sample, although in some cases both members of an interacting pair are observed by MaNGA. More details of the framework are provided in \citet{Nev23}.

In this framework, major mergers correspond to systems with merger stellar mass ratios $\mu \geq 1/4$, while minor mergers correspond to systems with $\mu < 1/4$. These major/minor classifications describe the inferred mass-ratio class of the merger system and do not require us to identify the companion galaxy or determine whether the MaNGA target is the more or less massive member of the pair. Merger stages are divided into early-stage, late-stage, and post-coalescence systems. Following \citet{Nev23} and \citet{Com24}, early-stage mergers are defined as systems roughly around the first pericentric passage and the average separation of the two galaxies is less than 10 kpc \citep{Nev19}\footnote{We emphasize that the definition of early-stage mergers adopted in this work, following \citet{Nev23} and \citet{Com24}, differs from that used in our previous studies \citep{Pan19,Pan25} and our ongoing work (Pan et al. submitted). In those studies, the early stage referred to the pre-encounter phase, when galaxies were still approaching each other before the first close passage and showed no clear signs of interaction-driven perturbations. In contrast, the early-stage systems in this work correspond to a more advanced phase around the first pericentric passage, where tidal interactions and merger-driven perturbations are already present.}, a phase typically associated with the first episode of tidally induced gas inflow. Late-stage mergers are defined as systems between apocentric passage and final coalescence, during which the galaxies re-approach and may undergo further gas inflows prior to the final coalescence. Post-coalescence mergers correspond to merger remnants from coalescence until 1~Gyr afterward. In this work, we adopt the 1~Gyr post-coalescence classification, following \citet{Nev23}, who showed that simulated galaxies can retain disturbed morphologies for up to $\sim$2.5~Gyr after a merger \citep{Bic21}.

The MaNGA Galaxy Merger Catalog has been used to investigate the connection between galaxy mergers and AGN activity \citep{Com24,Neg24}.  It has also been used to explore how galaxy mergers affect the stellar and gas kinematics of galaxies, as well as their dynamical evolution \citep{Rai25,Fu25}.

As our default classification, we adopt the hierarchical scheme described in \citet{Com24}. At the top level, we use the catalog columns \texttt{p\_merg\_stat\_50\_major\_merger} and \texttt{p\_merg\_stat\_50\_minor\_merger}, which represent the total probabilities of being a major or minor merger, respectively. Following \citet{Com24}, galaxies with both probabilities below 0.5 are classified as non-mergers, while systems with either probability greater than or equal to 0.5 are classified as mergers. 

For galaxies identified as mergers, we assign the merger type according to the larger of the two probabilities: systems with
 \path{p_merg_stat_50_major_merger} $>$  \path{p_merg_stat_50_major_minor}
are classified as major mergers, while those with
\path{p_merg_stat_50_major_merger} $<$  \path{p_merg_stat_50_major_minor}
are classified as minor mergers.

We further classify mergers into evolutionary stages using the stage-specific probabilities associated with the assigned merger type. The merger stage is determined by the highest probability among the early-stage, late-stage, and post-coalescence classes. For post-coalescence systems, the \emph{major} and \emph{minor} labels do not refer to directly observed galaxy pairs, but to the progenitor mass ratio inferred from the machine-learning classification  trained on merger simulations by \citet{Nev23}.

\subsection{Sample Selection}
We match the \texttt{Pipe3D} value-added data products with the MaNGA Galaxy Merger Catalog and obtain an initial sample of 7,655 galaxies. This is smaller than the full MaNGA sample ($\sim$10,000 galaxies) because the merger catalog does not include all MaNGA targets, primarily due to image quality limitations, failures in deriving reliable morphological measurements, or insufficient structural information required for robust merger classification \citep[see][]{Nev23}.

We further restrict the sample to galaxies with well-defined merger classifications. Systems whose probability combinations do not satisfy the mapping logic used to assign merger type and evolutionary stage (Section \ref{sec_merger_classification}) are excluded, reducing the sample to 7,499 galaxies.

Since reliable SFRs and gas-phase metallicities can only be measured in galaxies with sufficient star-forming spaxels, we further limit our analysis to star-forming systems. We adopt a global specific star formation rate cut of $\log(\mathrm{sSFR}/\mathrm{yr}^{-1}) > -11.0$, where the specific SFR (sSFR) is defined as SFR/\Mstar. After applying this selection, the final sample consists of 3,505 galaxies, including 1,745 non-mergers, 1,347 star-forming pre-coalescence mergers (regardless of merger type or stage) and 413 star-forming post-coalescence systems (regardless of merger type).

\begin{figure}
	\centering
	\includegraphics[width=0.45\textwidth]{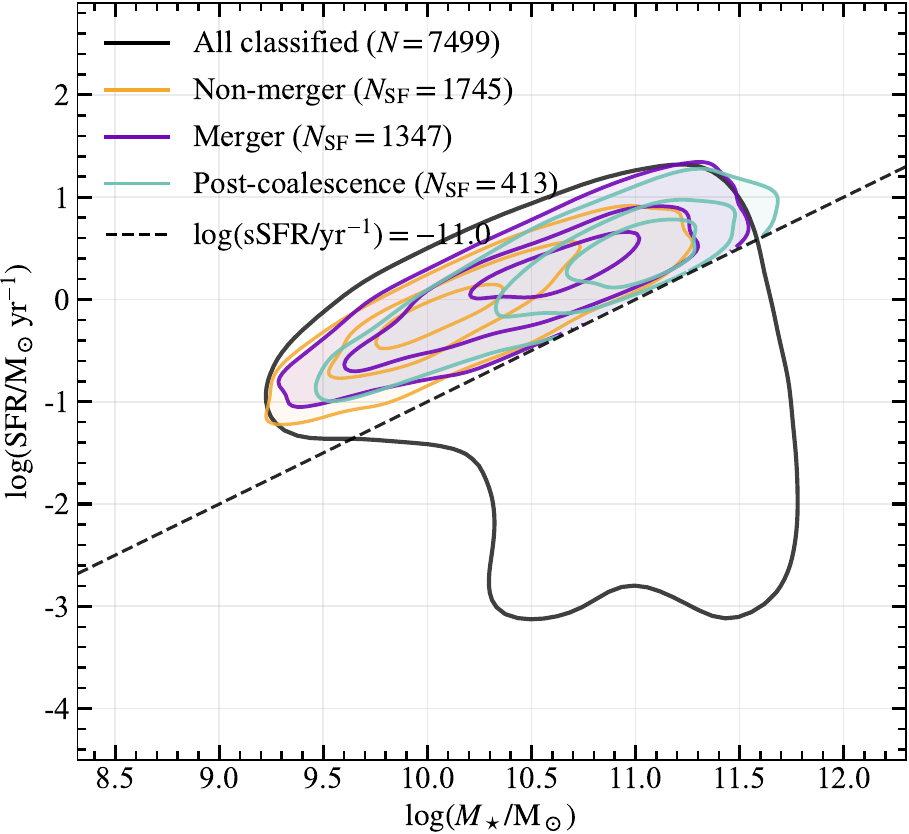}
	\caption{Distribution of the galaxy sample on the global SFR-\Mstar\ plane. The black contour shows the full sample of galaxies with valid merger classifications and marks the 20\% peak-density level to illustrate the overall distribution. The yellow, purple, and green contours represent the star-forming non-merger, pre-coalescence merger, and post-coalescence merger populations, respectively, with contour levels corresponding to 20\%, 50\%, and 80\% of the peak density in each subsample. The dashed line marks the adopted star-forming selection threshold of $\log(\mathrm{sSFR}/\mathrm{yr}^{-1}) = -11.0$. Both pre- and post-coalescence merger systems tend to occupy systematically higher \Mstar\ than the non-merger population. We therefore construct matched control samples to isolate the effects of mergers on star formation and  metallicity.
	}
	\label{fig_sample}
\end{figure}

Figure~\ref{fig_sample} shows the distribution of the sample on the global SFR--\Mstar\ plane. We find that merger and post-coalescence systems tend to occupy systematically higher \Mstar\ compared to non-mergers. This may suggest that mergers are more common among massive systems and/or may contribute to stellar mass growth through the assembly of progenitor galaxies and merger-driven star formation \citep{Bun09,Man16}. Therefore, to isolate the impact of mergers themselves, we construct matched control samples based on both global and local galaxy properties, as described in the next section.

\subsection{Control Sample Matching}
\label{sec_control_matching}

To isolate the impact of galaxy mergers on star formation and metallicity, we construct matched control samples at both the galaxy and spaxel levels. For each galaxy, control galaxies are selected from the non-merger population by matching in global \Mstar, redshift, \Reff, and S\'{e}rsic index, within tolerances of 0.1 dex in $\log M_\star$, 0.005 in redshift, 20\% in \Reff, and 1.0 in S\'{e}rsic index. This ensures that the control galaxies have similar global properties and observational conditions. In particular, matching in S\'{e}rsic index helps avoid comparing bulge-dominated and disk-dominated systems, which may exhibit systematically different star formation and metallicity properties driven by secular processes rather than mergers.

At the resolved level, each spaxel in a  galaxy is further matched to spaxels from the corresponding control galaxies by requiring similar galactocentric radius and local stellar mass surface density, within 0.2\,$R_{\rm e}$ and 0.1 dex in $\log \Sigma_\star$, respectively. This additional local matching minimizes biases arising from radial gradients and local environmental differences within galaxies.

For each spaxel, the median value of the matched control spaxels is adopted as the reference. The local enhancement (or suppression) is then defined as the logarithmic difference between the spaxel in question and its matched control median, denoted as \Dsf\ and \Doh\ for star formation and metallicity, respectively. Since these quantities are measured in logarithmic units, positive values indicate enhancement relative to the control sample, while negative values indicate suppression. This method allows us to quantify the impact of mergers on ISM properties while controlling for both global galaxy-scale and local resolved-scale dependencies. The same matching procedure is also applied to non-merger galaxies. The resulting distributions of \Dsf\ and \Doh\ for non-mergers provide reference distributions for the offsets measured in galaxies classified as non-mergers. In Section \ref{sec_result}, we quantify the differences between the merger and non-merger populations using the median offsets and their 95\% galaxy-bootstrap confidence intervals, together with Anderson--Darling tests comparing the full distributions. This control-matching approach has been widely adopted in studies of merger-driven activity, including variations in star formation and metallicity, changes in gas properties, kinematic asymmetry, and AGN triggering \citep[e.g.,][]{Ell11a,Scu12,Vio18,Ell18a,Ell18b,Pan19,Fen20,Tho22,Byr23,Pan25,Shi26}, as well as in studies of bar-driven processes \citep[e.g.,][]{Ell11b,Cha26}.

Although this work adopts a specific metallicity calibration (\emph{O3N2}), we have shown that the metallicity offsets, both in the radial profiles \citep{Pan25} and in the fitted gradients (Pan et al. submitted), remain consistent when using alternative calibrations. 
Different calibrations may introduce systematic shifts in the absolute metallicity scale \citep[e.g.,][]{Scu21}, but these have little impact on the relative offsets considered here.

\section{Results}
\label{sec_result}
\subsection{Merger-induced Variations in Star Formation and Chemical Enrichment}
\label{sec_result_spx_dist}

To investigate how galaxy mergers affect star formation and gas-phase metallicity on resolved scales, we first examine the spaxel-by-spaxel distributions of offsets \Dsf\ and \Doh. By comparing the full distributions of these offsets within 1.5~\Reff, we quantify the overall prevalence, strength, and diversity of merger-induced star formation enhancement and metallicity dilution, independent of their spatial location.

\subsubsection{Resolved Star Formation Rate Offsets}
\label{sec_sfr_spx}

\begin{figure*}[!ht]
	\centering
	\subfigure[]{\label{fig_dist_ssfr_early}\includegraphics[width=0.9\textwidth]{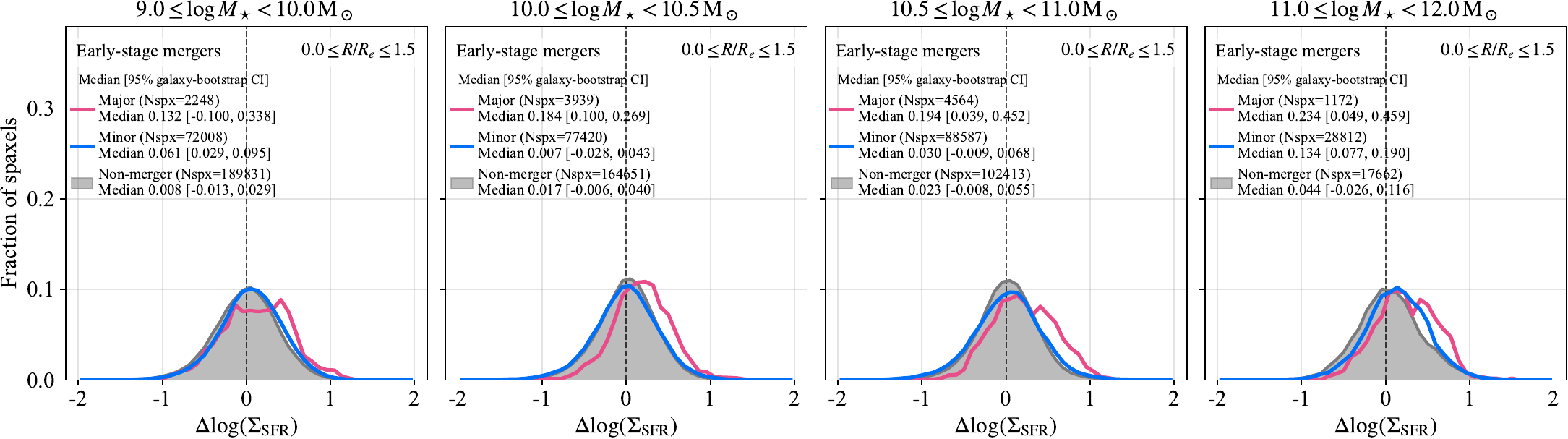}}
	\hspace{8pt}
	\subfigure[]{\label{fig_dist_ssfr_late}\includegraphics[width=0.9\textwidth]{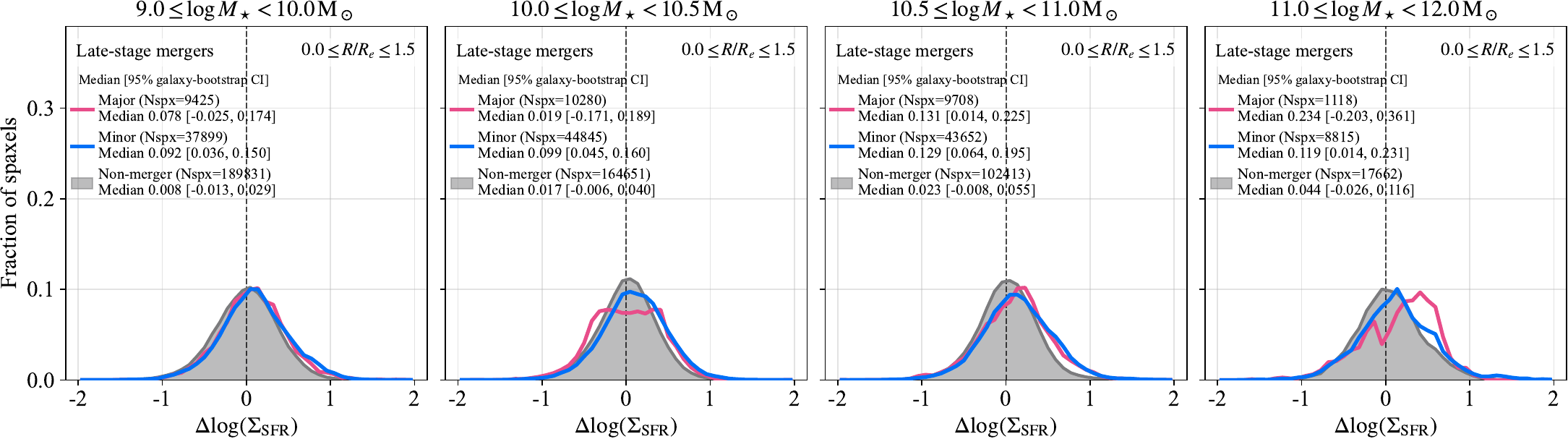}}
	\subfigure[]{\label{fig_dist_ssfr_post}\includegraphics[width=0.9\textwidth]{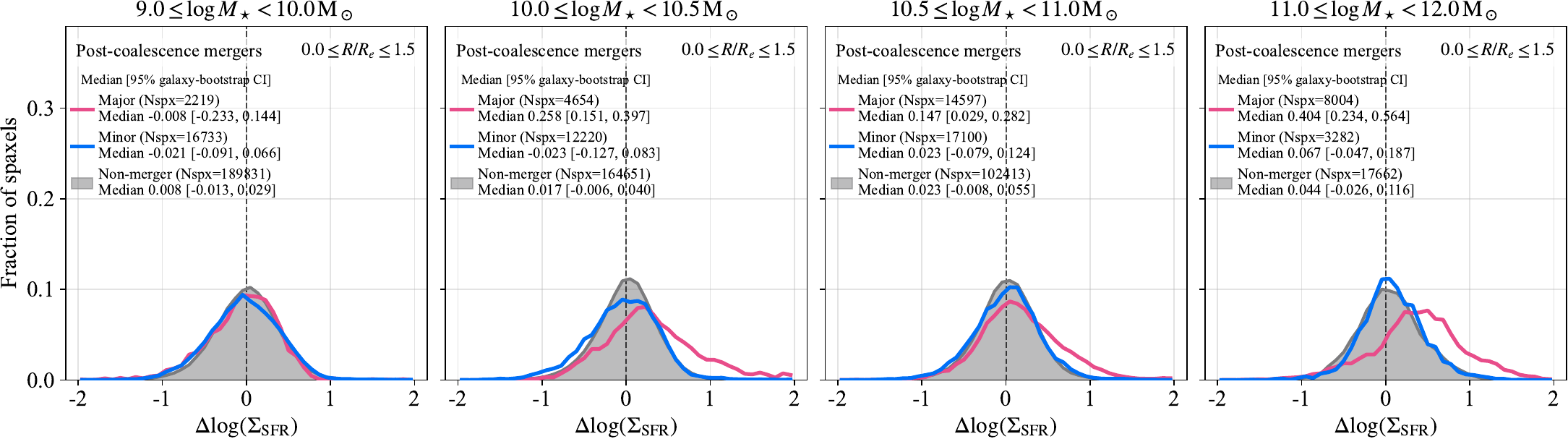}}

	\caption{Normalized distributions of resolved star formation rate surface density offsets, \Dsf, relative to matched non-merger controls for spaxels within 1.5~\Reff. The three rows show early-stage ({\bf top}), late-stage ({\bf middle}), and post-coalescence ({\bf bottom}) mergers, while the four columns from {\bf left} to {\bf right} represent increasing \Mstar. Major mergers are shown in red, minor mergers in blue, and the non-merger control sample in gray. The vertical dashed line marks \Dsf\ $= 0$, corresponding to no offset relative to the control sample. The median \Dsf\ and corresponding 95\% galaxy-bootstrap confidence interval (CI) for each population are reported in each panel. Major mergers generally exhibit  stronger positive offsets than minor mergers, particularly in the early-stage and post-coalescence phases, indicating stronger merger-driven star formation enhancement. In contrast, minor mergers show only modest star formation enhancement in terms of their median \Dsf\ across most merger stages and stellar-mass bins.
	}
	\label{fig_dist_ssfr}
\end{figure*}

\begin{figure*}[!ht]
	\centering
	\subfigure[]{\label{fig_sf_stat_ad}\includegraphics[width=0.45\textwidth]{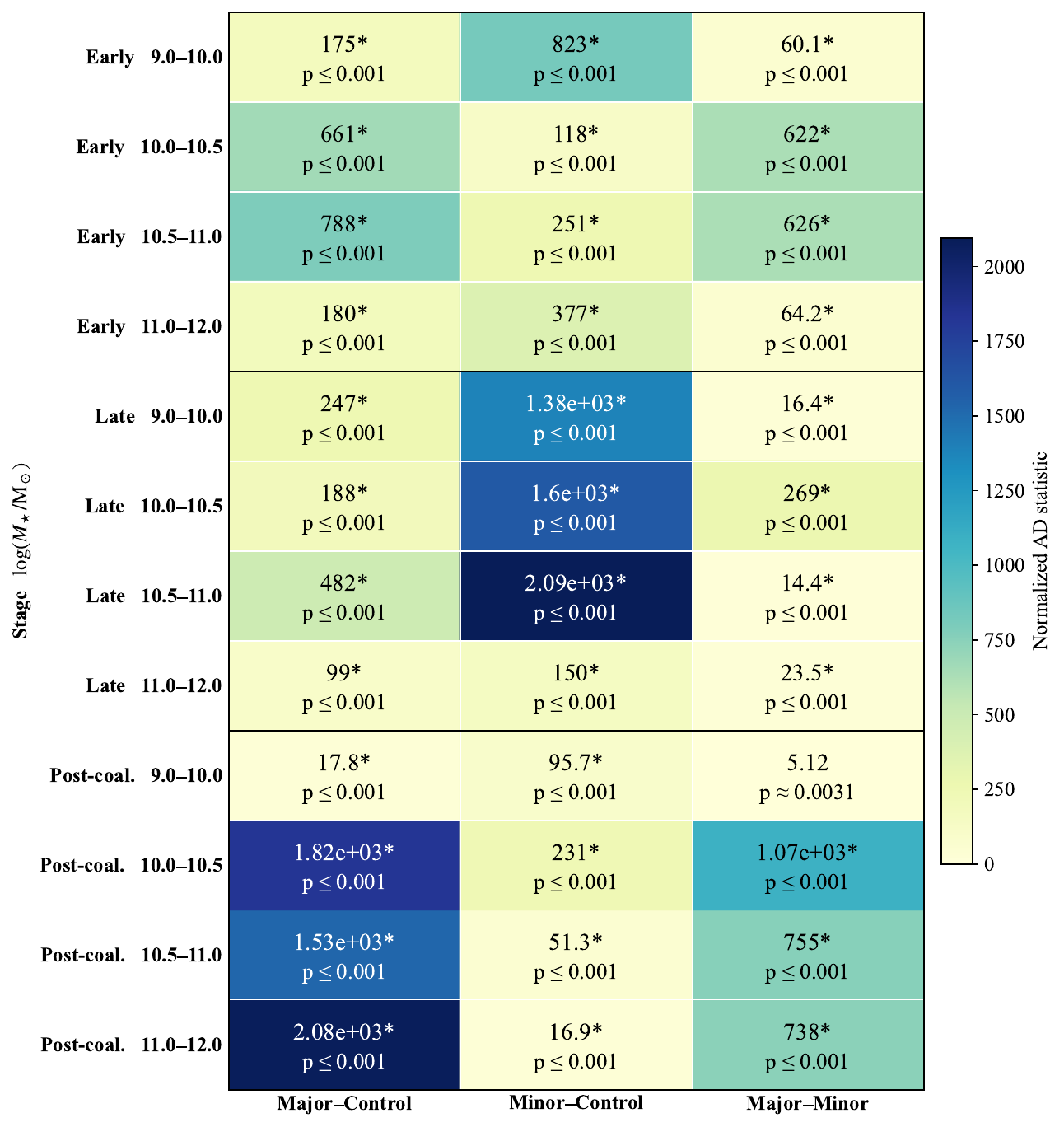}}
	\hspace{8pt}
	\subfigure[]{\label{fig_oh_stat_ad}\includegraphics[width=0.45\textwidth]{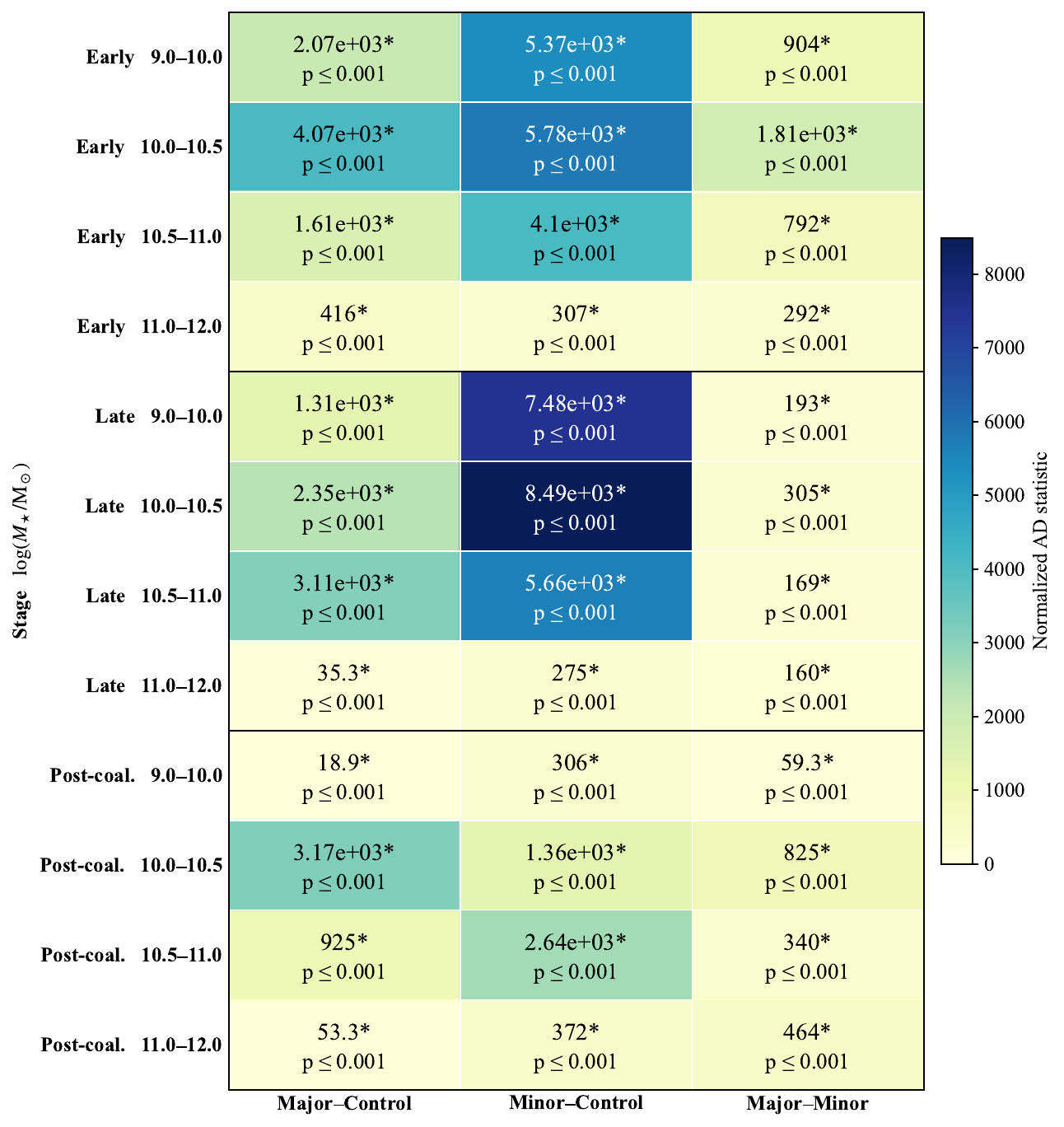}}
	
\caption{Anderson--Darling (AD) two-sample test statistics for the resolved star formation rate surface density ({\bf panel (a)}) and gas-phase metallicity ({\bf panel (b)}) offset distributions. Results are shown separately for the early, late, and post-coalescence merger stages and in four stellar-mass bins. The three columns compare the Major--Control, Minor--Control, and Major--Minor distributions, respectively.  The corresponding $p$-value is reported below the AD statistic in each cell. Asterisks indicate statistically significant differences between the two distributions according to our adopted criterion of $p<0.003$.}
	\label{fig_stat_ad}
\end{figure*}

Figure \ref{fig_dist_ssfr} presents the normalized distributions of \Dsf, separated by merger stage (top to bottom) and \Mstar\ (left to right), for spaxels within 1.5~\Reff. The three rows correspond to early-stage, late-stage, 
and post-coalescence mergers, respectively, while the four columns represent 
different \Mstar\ bins increasing from left to right. Major mergers are shown 
in red, minor mergers in blue, and the control non-merger population in gray.
To quantify the characteristic offsets, Figure \ref{fig_dist_ssfr} also reports the median \Dsf\ and its 95\% galaxy-bootstrap confidence interval for each population, estimated using 2,000 bootstrap resamples. We further compare the full distributions using two-sample Anderson--Darling (AD) tests (Figure \ref{fig_sf_stat_ad}), adopting $p<0.003$ as the criterion for statistical significance, corresponding approximately to the conventional $3\sigma$ threshold.
Since the large number of spaxels can make even small distributional differences statistically significant, we consider the AD statistics together with the median offsets when assessing the magnitude of the merger-related effects.

For early-stage mergers (Figure \ref{fig_dist_ssfr_early}), minor mergers show modest positive median \Dsf\ values, ranging from $\sim0.007$ to $0.13$ dex across the four \Mstar\ bins, indicating modest star formation enhancement relative to their matched controls. Major mergers exhibit systematically larger positive median \Dsf\ values of $\sim0.13$--$0.23$ dex, indicating stronger star formation enhancement. 
For both merger types, the largest median \Dsf\ is found in the highest-\Mstar\ bin.
The control medians remain close to zero in all four \Mstar\ bins.
The AD tests show statistically significant differences for the Major--Control, Minor--Control, and Major--Minor distributions in all four \Mstar\ bins ($p\leq0.001$; Figure
\ref{fig_sf_stat_ad}).

For late-stage mergers (Figure \ref{fig_dist_ssfr_late}), minor mergers show modest but relatively constant positive median \Dsf\ values of $\sim$0.05--0.1 dex across the stellar-mass bins. These median enhancements are generally comparable to or larger than those of their early-stage minor-merger counterparts, but remain modest compared with those observed in early-stage major mergers. Major mergers show a stronger dependence on \Mstar, with relatively small median \Dsf\ values in the two lower-mass bins ($0.08$ and $0.02$ dex) and larger values in the two higher-mass bins ($0.13$ and $0.23$ dex). Compared with early-stage major mergers, the median enhancement is weaker in the three lower-\Mstar\ bins but similar in the highest-\Mstar\ bin. The differences between the major- and minor-merger median offsets are also generally smaller at the late stage than at the early stage. Nevertheless, the AD tests show statistically significant differences among the Major--Control, Minor--Control, and Major--Minor distributions in all four \Mstar\ bins ($p\leq0.001$; Figure \ref{fig_sf_stat_ad}).

The most distinct differences emerge in post-coalescence systems
(Figure \ref{fig_dist_ssfr_post}).
For major mergers, the median \Dsf\ is close to zero in the lowest-\Mstar\ bin ($-0.008$ dex), indicating little change in star formation relative to the matched controls. In contrast, the three higher-\Mstar\ bins show positive median \Dsf\ values of $0.26$, $0.15$, and $0.40$ dex, respectively, indicating substantial star formation enhancement, with the largest median enhancement occurring in the highest-\Mstar\ bin. The major-merger distributions at intermediate and high stellar masses also show extended positive tails, reaching \Dsf\ $\sim1$ dex and beyond. The distribution-shape statistics reported in Appendix \ref{appendix_sf} and Figure \ref{fig_sf_stat_all} further quantify these differences in distribution shape and confirm their statistical significance.
Minor mergers show much smaller median offsets, with \Dsf\ ranging from $-0.02$ to $0.07$ dex across the four \Mstar\ bins, indicating comparatively weak changes in star formation. The AD tests  show statistically significant Major--Control and Minor--Control distributional differences in all four \Mstar\ bins ($p\leq0.001$; Figure \ref{fig_sf_stat_ad}). The Major--Minor differences are also
significant in the three higher-\Mstar\ bins ($p\leq0.001$), whereas the lowest-\Mstar\ comparison has $p=0.0031$, marginally above our adopted $p<0.003$ significance criterion.

Overall, the median \Dsf\ measurements quantitatively support a stronger star formation response in major than in minor mergers, particularly in the early-stage and post-coalescence phases. Major mergers show positive median \Dsf\ across most merger stages and \Mstar\ bins, with particularly strong enhancements in intermediate- and high-\Mstar\ post-coalescence systems, whereas the median offsets of minor mergers generally remain modest. The AD tests
further show that the full \Dsf\ distributions of the merger
populations are statistically distinguishable from the controls in nearly all cases, even where the differences in median \Dsf\ are small. The additional skewness, kurtosis, and normality statistics presented in Appendix \ref{appendix_sf} and Figure \ref{fig_sf_stat_all} further characterize differences in the shapes of these distributions.

\subsubsection{Resolved Gas-phase Metallicity Offsets}
\label{sec_oh_spx}

\begin{figure*}[!ht]
	\centering
	\subfigure[]{\label{fig_dist_oh_early}\includegraphics[width=0.9\textwidth]{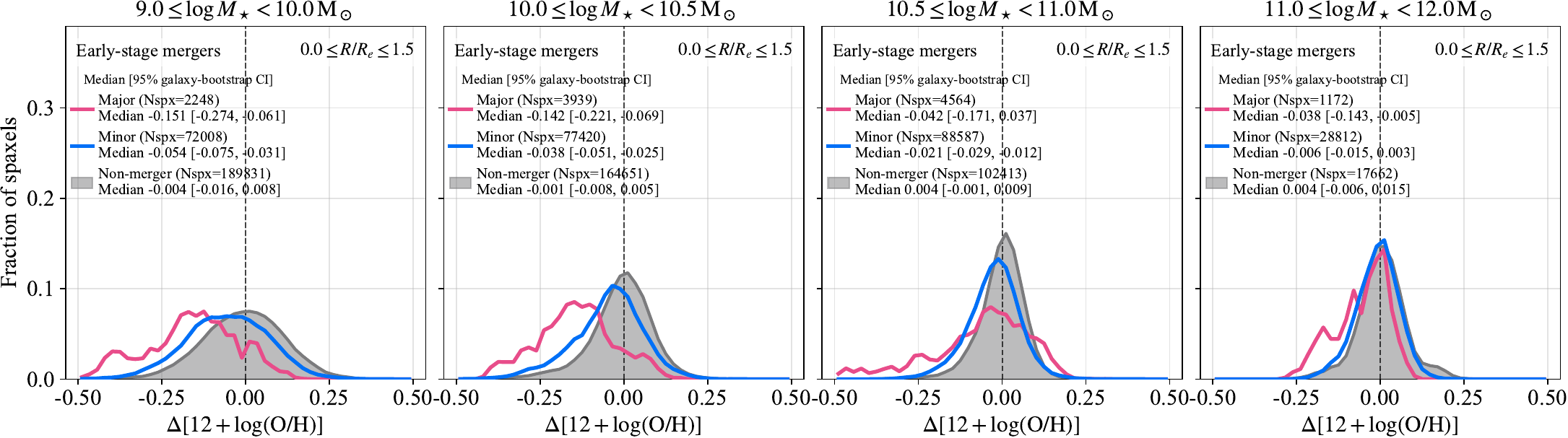}}
	\hspace{8pt}
	\subfigure[]{\label{fig_dist_oh_late}\includegraphics[width=0.9\textwidth]{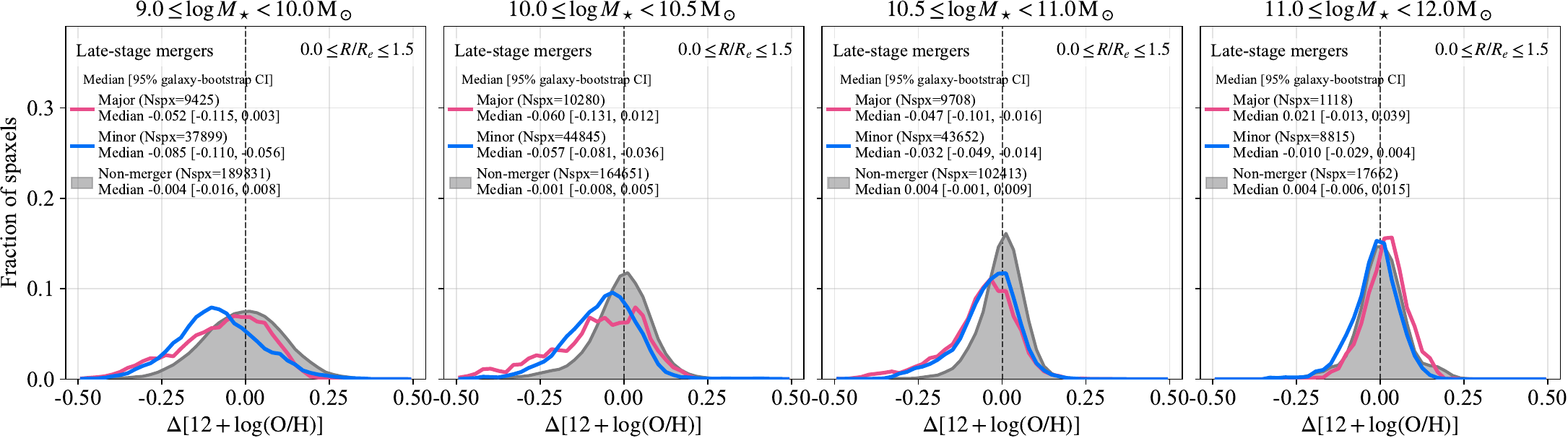}}
	\subfigure[]{\label{fig_dist_oh_post}\includegraphics[width=0.9\textwidth]{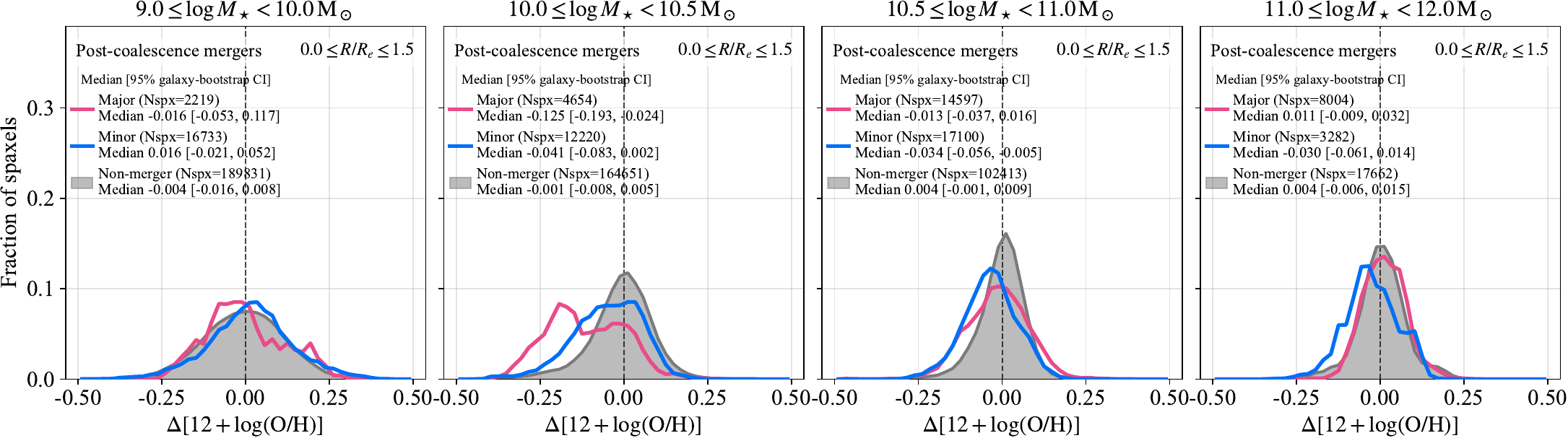}}
	
	\caption{Normalized distributions of resolved gas-phase metallicity offsets, \Doh, relative to matched non-merger controls for spaxels within 1.5~\Reff. The plotting style and panel layout are the same as in Figure~\ref{fig_dist_ssfr}. The vertical dashed line marks \Doh\ $= 0$, corresponding to no offset relative to the control sample. The median \Doh\ and corresponding 95\% galaxy-bootstrap confidence interval (CI) for each population are reported in each panel. In general, metallicity dilution is stronger in lower-mass than in higher-mass systems for both merger types and is more pronounced in the early and late stages than in post-coalescence systems.
	%Major mergers generally exhibit broader distributions and stronger negative offsets than minor mergers, particularly in the early-stage and lower-\Mstar\ bins, indicating stronger merger-driven metallicity dilution. In contrast, minor mergers show weaker negative offsets.
	} 
	\label{fig_dist_oh}
\end{figure*}

Figure \ref{fig_dist_oh} presents the normalized distributions of resolved gas-phase metallicity offsets, \Doh, with the same panel layout as Figure \ref{fig_dist_ssfr}.
The median \Doh\ and its 95\% galaxy-bootstrap confidence interval are reported for each population in Figure \ref{fig_dist_oh}. We also quantify differences between the full distributions using two-sample AD tests (Figure \ref{fig_oh_stat_ad}). Additional distribution-shape statistics are presented in Appendix \ref{appendix_oh} and Figure \ref{fig_oh_stat_all}.

For early-stage mergers (Figure \ref{fig_dist_oh_early}), major mergers show negative median \Doh\ in all four \Mstar\ bins, indicating metallicity dilution relative to their matched controls. The dilution is strongest in the two lowest-\Mstar\ bins, with median \Doh\ values of $-0.15$ and $-0.14$ dex, and becomes substantially weaker at higher stellar masses, with median values of $\sim$ $-0.04$ dex. Nevertheless, extended tails toward low \Doh\ are present across all four \Mstar\ bins. Minor mergers also show negative median offsets, but with smaller magnitudes and a clear dependence on \Mstar. The median dilution progressively decreases from $\sim$ $-0.05$ to $-0.04$, $-0.02$, and $-0.006$ dex with increasing \Mstar. Thus, early-stage metallicity dilution is stronger in major than in minor mergers and is most pronounced at lower stellar masses. The AD tests show statistically significant differences among the Major--Control, Minor--Control, and Major--Minor distributions in all four \Mstar\ bins ($p\leq0.001$; Figure \ref{fig_oh_stat_ad}).

For late-stage mergers (Figure \ref{fig_dist_oh_late}), negative median \Doh\ values persist for both major and minor mergers in most \Mstar\ bins. Major mergers have median \Doh\ values of $-0.05$, $-0.06$, $-0.05$, and $+0.02$ dex with increasing \Mstar, while the corresponding minor-merger medians are $-0.08$, $-0.06$, $-0.03$, and $-0.01$ dex. The offsets therefore remain predominantly negative below $\log(M_\star/M_\odot)<11$ for both merger types, whereas their magnitudes are small and close to zero in the highest-\Mstar\ bin. Interestingly, the median offsets show different changes from the early to late stage for major and minor mergers. For major mergers, the median \Doh\ values in the two lowest-\Mstar\ bins become less negative, changing from $-0.15$ and $-0.14$ dex in the early stage to $-0.05$ and $-0.06$ dex in the late stage, respectively. In contrast, minor mergers show more negative median offsets in the late stage in the three lower-\Mstar\ bins, with the largest change occurring in the lowest-\Mstar\ bin.  Thus, the median \Doh\ measurements suggest different stage-dependent behavior for major and minor mergers, although the 95\% galaxy-bootstrap confidence intervals overlap between the early- and late-stage populations. Stronger metallicity dilution nevertheless remains generally associated with lower stellar masses, regardless of merger type. The AD tests show statistically significant differences among the Major--Control, Minor--Control, and Major--Minor distributions in all four \Mstar\ bins ($p\leq0.001$; Figure \ref{fig_oh_stat_ad}).

In post-coalescence systems (Figure \ref{fig_dist_oh_post}), the median metallicity offsets are generally small. For both merger types, the median $|$\Doh$|$ is less than $0.04$ dex in all \Mstar\ bins, except for major mergers at $10.0\leq\log(M_\star/M_\odot)<10.5$, where a substantial negative offset persists (\Doh$\sim-0.12$ dex). Thus, the post-coalescence populations generally show smaller median metallicity offsets than those observed at earlier merger stages. Despite the generally small median offsets, the distributions retain extended tails toward low \Doh\ or show departures from a typical Gaussian distribution in some \Mstar\ bins for both merger types. These distribution shapes are further characterized and their statistical significance is assessed in Appendix \ref{appendix_oh} and Figure \ref{fig_oh_stat_all}. The AD tests show statistically significant differences for all Major--Control, Minor--Control, and Major--Minor comparisons ($p\leq0.001$), indicating detectable differences in the full distributions even where the median offsets are small.

Overall, the median \Doh\ measurements show that the strongest metallicity dilution occurs in early-stage major mergers at low stellar masses. The magnitude of the dilution generally decreases toward higher \Mstar. In post-coalescence systems, the median metallicity offsets are generally small for both major and minor mergers. However, the AD tests demonstrate statistically significant differences among the full metallicity-offset distributions in all population comparisons, including cases with small median offsets. Additional skewness, kurtosis, and normality statistics characterizing the shapes of these distributions are presented in Appendix \ref{appendix_oh} and Figure \ref{fig_oh_stat_all}.

\subsection{Radial Profiles of Star Formation and Metallicity Offsets}
\label{sec_result_radial}
To further investigate where merger-driven perturbations occur within galaxies, we examine the radial profiles of the resolved star formation rate and gas-phase metallicity offsets, \Dsf\ and \Doh. While the previous section focused on the overall distributions of these quantities, radial profiles provide a complementary view by tracing their average behavior as a function of galactocentric radius. This allows us to assess whether star formation enhancement and metallicity dilution are centrally concentrated or extend to larger galactic radii, and how these trends vary with merger stage and \Mstar. In addition to the binned radial profiles, we also use Kendall's rank correlation test on the original unbinned spaxel measurements to quantify monotonic radial trends independently of the adopted radial binning.

\subsubsection{Radial Profiles of Star Formation Rate Offsets}
\label{sec_sfr_rad}

\begin{figure*}[!ht]
	\centering
	\subfigure[]{\label{fig_rad_ssfr_early}\includegraphics[width=0.9\textwidth]{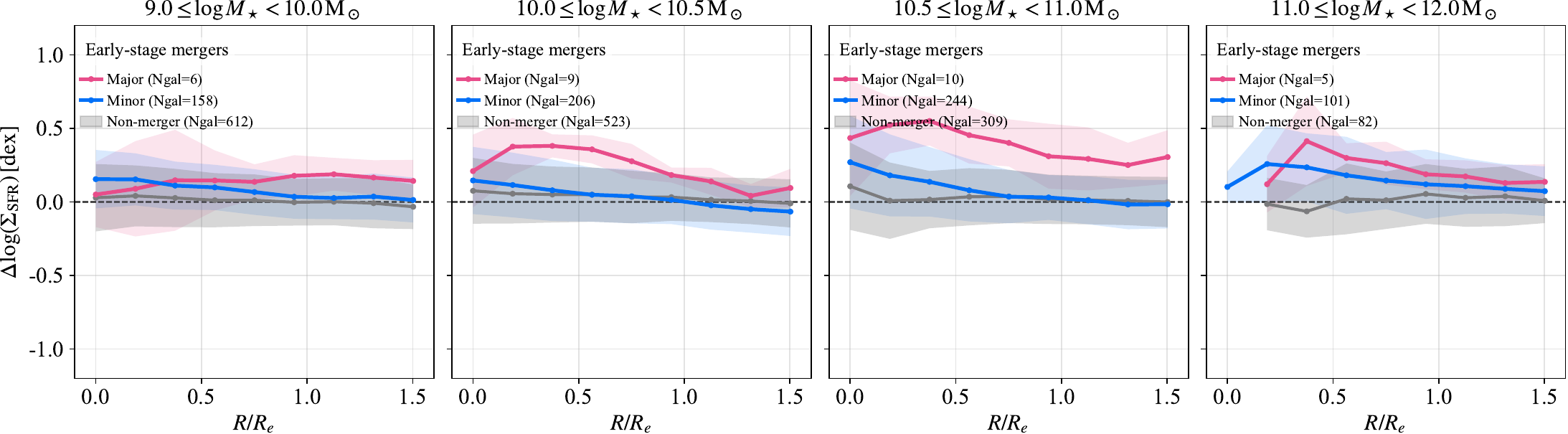}}
	\hspace{8pt}
	\subfigure[]{\label{fig_rad_ssfr_late}\includegraphics[width=0.9\textwidth]{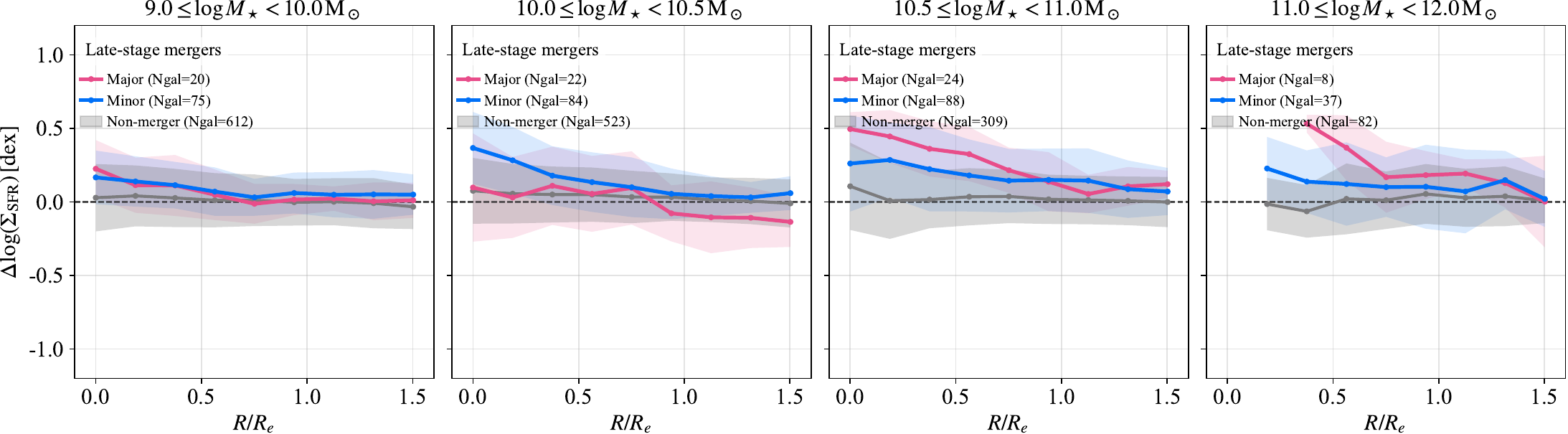}}
	\subfigure[]{\label{fig_rad_ssfr_post}\includegraphics[width=0.9\textwidth]{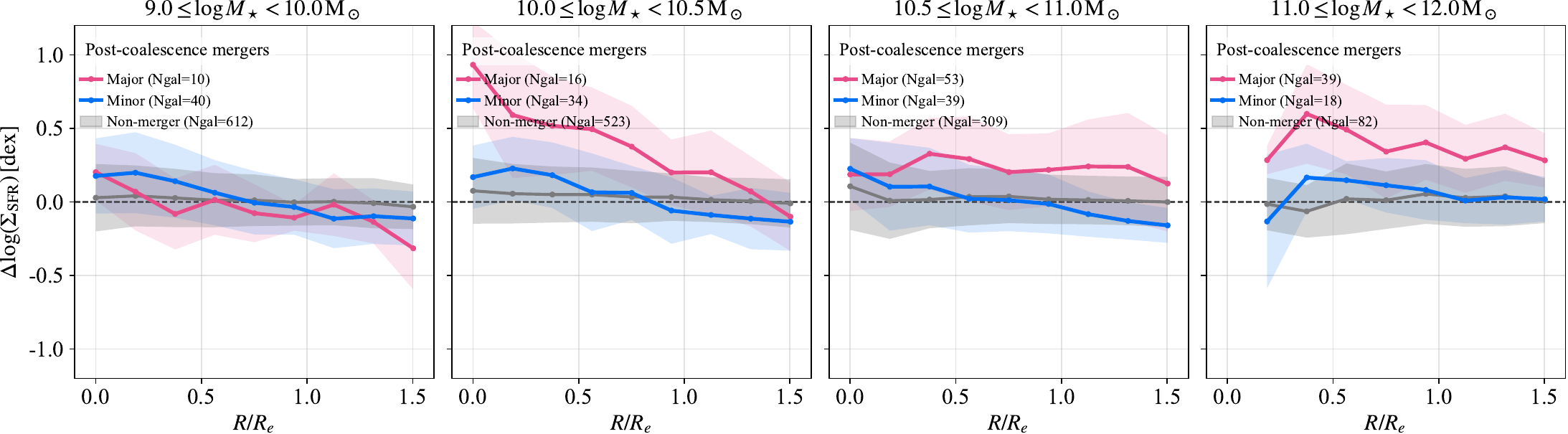}}
	
	\caption{Median radial profiles of resolved star formation rate surface density offsets, \Dsf, as a function of galactocentric radius normalized by the effective radius, $R/$\Reff. The plotting style and panel layout are the same as in Figure~\ref{fig_dist_ssfr}. Major mergers are shown in red, minor mergers in blue, and the non-merger control sample in gray. The shaded regions show the median absolute deviation (MAD) across galaxies in each radial bin, representing galaxy-to-galaxy scatter rather than the uncertainty on the median. Major mergers generally exhibit stronger and more spatially extended positive \Dsf\ than minor mergers, particularly in the early-stage and post-coalescence phases. In contrast, minor mergers show weaker and more centrally concentrated star formation enhancement.
	}
	\label{fig_rad_ssfr}
\end{figure*}

Figure \ref{fig_rad_ssfr} presents the median radial profiles of \Dsf\ as a function of galactocentric radius normalized by the effective radius, $R$/\Reff. The profiles are separated by merger stage (rows) and \Mstar\ (columns), with major mergers shown in red, minor mergers in blue, and non-merger controls in gray. To construct the median radial profiles, we first  calculate the median \Dsf\ within each radial bin for each individual galaxy and then compute the median of these galaxy-level measurements across all galaxies within each merger stage and \Mstar\ bin. For an individual galaxy to contribute to a given radial bin, we require at least five valid spaxels in that bin. The $N_{\rm gal}$ values reported in Figure \ref{fig_rad_ssfr} give the total number of distinct galaxies contributing to the radial profile within each population and \Mstar\ bin. Since the minimum-spaxel criterion is applied independently to each radial bin, the number of galaxies contributing to an individual radial bin can vary with radius.
In addition, each stacked radial bin must contain contributions from at least three galaxies. Radial bins with fewer than three contributing galaxies are not shown. The shaded regions show the median absolute deviation (MAD) across the contributing galaxies in each radial bin and characterize the galaxy-to-galaxy dispersion about the median profiles rather than the uncertainty on the median.

For early-stage mergers (Figure \ref{fig_rad_ssfr_early}), major mergers exhibit systematically positive median \Dsf\ profiles, with enhancements of $\sim$$0.2$--$0.5$ dex across most \Mstar\ bins, indicating enhanced star formation relative to the control sample. The median profiles of major mergers are generally flat and spatially extended, indicating widespread star formation enhancement across the radial range explored. In contrast, minor mergers show more modest enhancement in their median profiles, reaching at most $\sim$$0.2$ dex in the central regions, and gradually declining with radius to become comparable to or slightly below the control sample in the outer regions. This indicates that the median star formation enhancement in major mergers is both stronger and more spatially extended, whereas in minor mergers it remains weaker and more centrally concentrated.

For late-stage mergers (Figure \ref{fig_rad_ssfr_late}), the median radial \Dsf\ profiles show a clear dependence on \Mstar\ and merger type. In the two lower-\Mstar\ bins, major mergers show relatively weak median enhancement ($<0.2$ dex) that decreases with radius and approaches zero or becomes slightly negative in the outer regions. In contrast, the two higher-\Mstar\ bins show stronger central enhancement ($\sim$ 0.5 dex), comparable to that seen in the corresponding early-stage median profiles, but with a clearer radial dependence, declining toward larger radii. Minor mergers also show positive median \Dsf\ in the central regions across the \Mstar\ bins, reaching up to $\sim0.25$ dex, with the enhancement generally decreasing toward larger radii. 

In post-coalescence systems (Figure~\ref{fig_rad_ssfr_post}), strong star formation enhancement is observed again in major mergers, demonstrating a clear dependence on merger stage. The positive \Dsf\ is most pronounced at intermediate radii ($R$/\Reff $\lesssim$ 1.0), particularly in the intermediate- and high-mass bins ($\log M_\star/M_\odot > 10.0$), with offsets reaching $\sim0.2$--$1.0$ dex. This stellar-mass dependence is also evident in the early- and late-stage major mergers, where intermediate- and high-mass systems tend to exhibit stronger enhancement than their lower-mass counterparts. Compared with the late-stage profiles, the post-coalescence major mergers at intermediate and high \Mstar\ generally show less pronounced radial variation, more closely resembling the early-stage profiles, with enhanced \Dsf\ extending over a broad radial range. In contrast, the lowest-mass post-coalescence major mergers remain close to zero over most radii, with no clear enhancement or suppression of star formation. Minor mergers exhibit only mild enhancement in the central regions, with \Dsf\ declining toward zero or becoming slightly suppressed at larger radii, similar to the trends seen in the other merger stages.

\begin{figure*}[!ht]
	\centering
\subfigure[]{\label{fig_rad_ssfr_stat_tau}\includegraphics[width=0.45\textwidth]{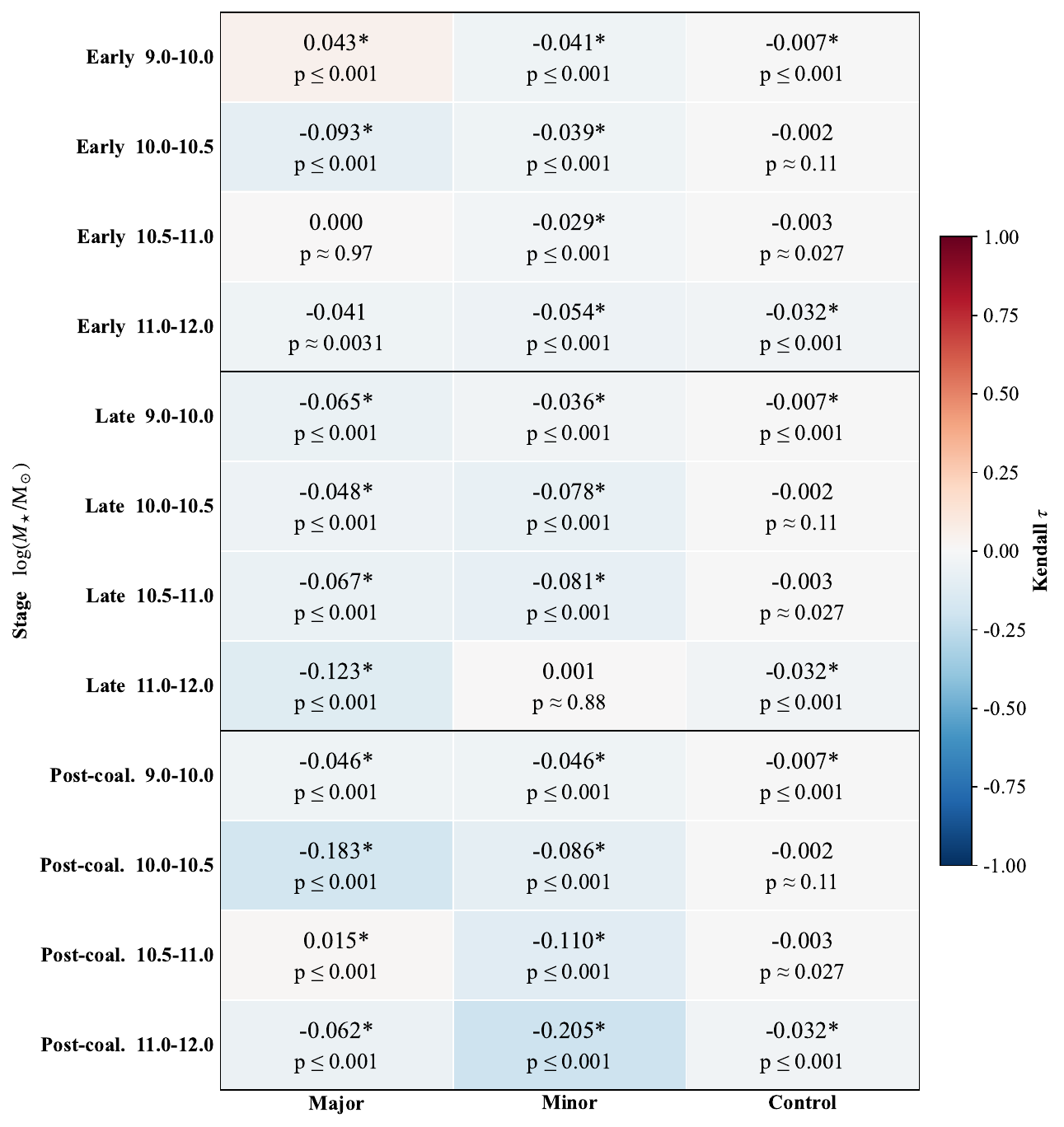}}
\hspace{8pt}
\subfigure[]{\label{fig_rad_oh_stat_tau}\includegraphics[width=0.45\textwidth]{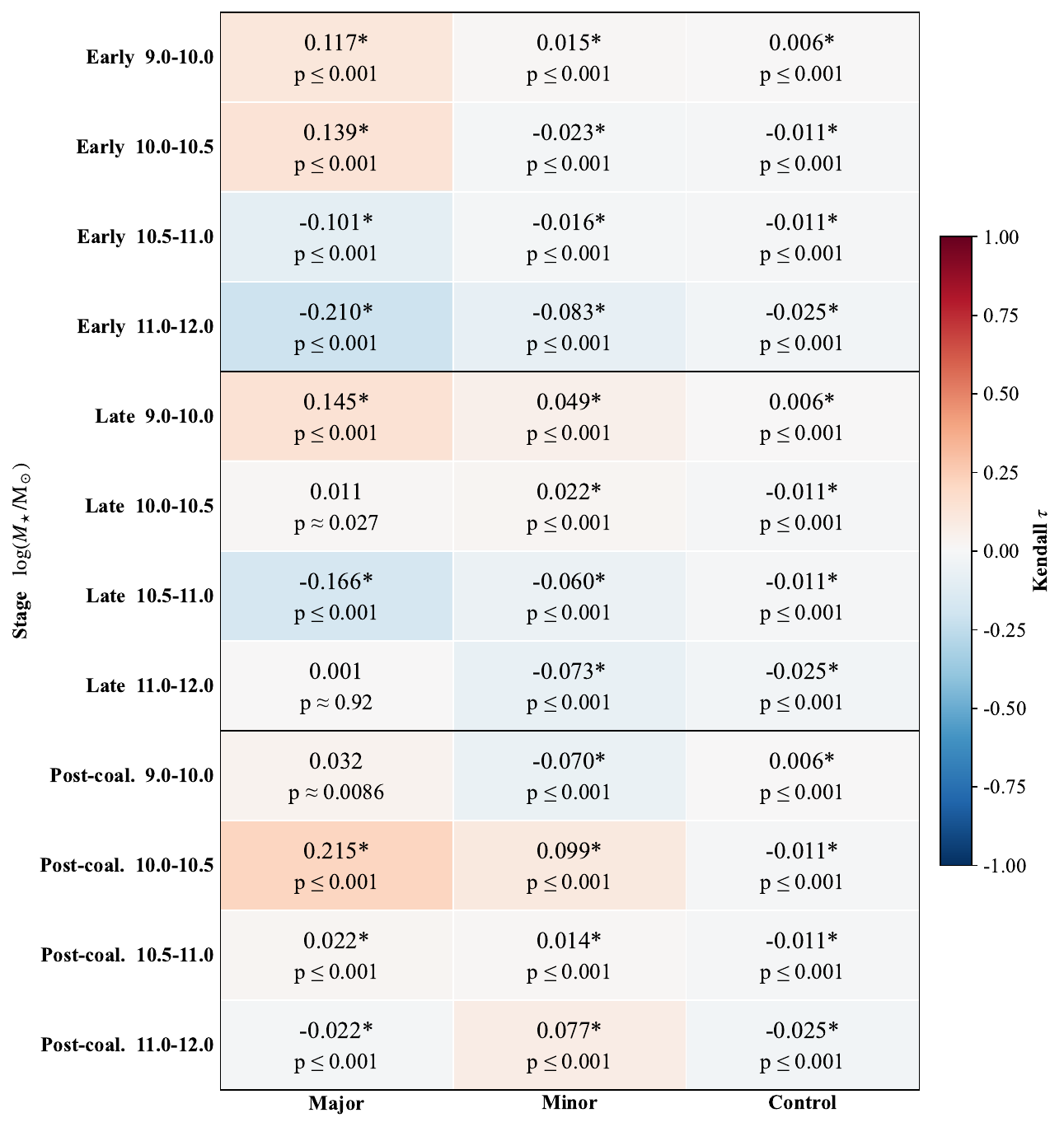}}
	\caption{Kendall rank-correlation statistics for the resolved star formation rate surface density ({\bf panel (a)}) and gas-phase metallicity ({\bf panel (b)}) offsets as a function of normalized galactocentric radius, $R$/\Reff, for major mergers, minor mergers, and their matched control galaxies. Results are shown separately for the early, late, and post-coalescence merger stages and in four stellar-mass bins. Positive (negative) $\tau$ values indicate that the corresponding offset tends to increase (decrease) with increasing radius. The corresponding $p$-values for the null hypothesis of no correlation are reported below each $\tau$ value. Asterisks indicate statistically significant correlations according to our adopted criterion of $p<0.003$.}
	\label{fig_rad_stat_tau}
\end{figure*}

To quantify these radial behaviors independently of the adopted radial binning, Figure \ref{fig_rad_ssfr_stat_tau} presents Kendall's rank correlation coefficients between \Dsf\ and $R$/\Reff, calculated using the original unbinned spaxel measurements. Many merger populations show statistically significant correlations according to our adopted criterion of $p<0.003$, although the associations are generally weak ($|\tau|\lesssim0.2$). Negative correlations, corresponding to decreasing \Dsf\ with radius, are particularly evident in late-stage major mergers ($\tau=-0.067$ to $-0.123$) and in some post-coalescence populations, including major mergers at $10.0\leq\log(M_\star/M_\odot)<10.5$ ($\tau=-0.183$) and minor mergers at $11.0\leq\log(M_\star/M_\odot)<12.0$ ($\tau=-0.205$), all with $p\leq0.001$. Thus, several merger populations show statistically significant monotonic radial trends in \Dsf, although the generally small $|\tau|$ values indicate that these trends are weak relative to the substantial spaxel-to-spaxel scatter.

Overall, Figure \ref{fig_rad_ssfr} reveals a clear dependence of star formation enhancement on merger stage, merger type, and \Mstar. Major mergers generally show stronger median \Dsf\ enhancement than minor mergers, particularly at intermediate and high \Mstar. The strongest and most spatially extended enhancements are generally seen in early-stage and post-coalescence major mergers, with typical \Dsf\ values of $\sim0.2$--$0.5$ dex extending across much of the radial range explored in this work. In contrast, the strongest enhancement in minor mergers is generally $\lesssim0.25$ dex and concentrated within $R$/\Reff$\lesssim0.5$, with \Dsf\ decreasing toward zero at larger radii. Thus, the resolved SFR response depends jointly on merger type, merger stage, stellar mass, and radius rather than following a uniform trend across the merger sequence.

\subsubsection{Radial Profiles of Gas-phase Metallicity Offsets}
\label{sec_oh_rad}

\begin{figure*}[!ht]
	\centering
	\subfigure[]{\label{fig_rad_oh_early}\includegraphics[width=0.9\textwidth]{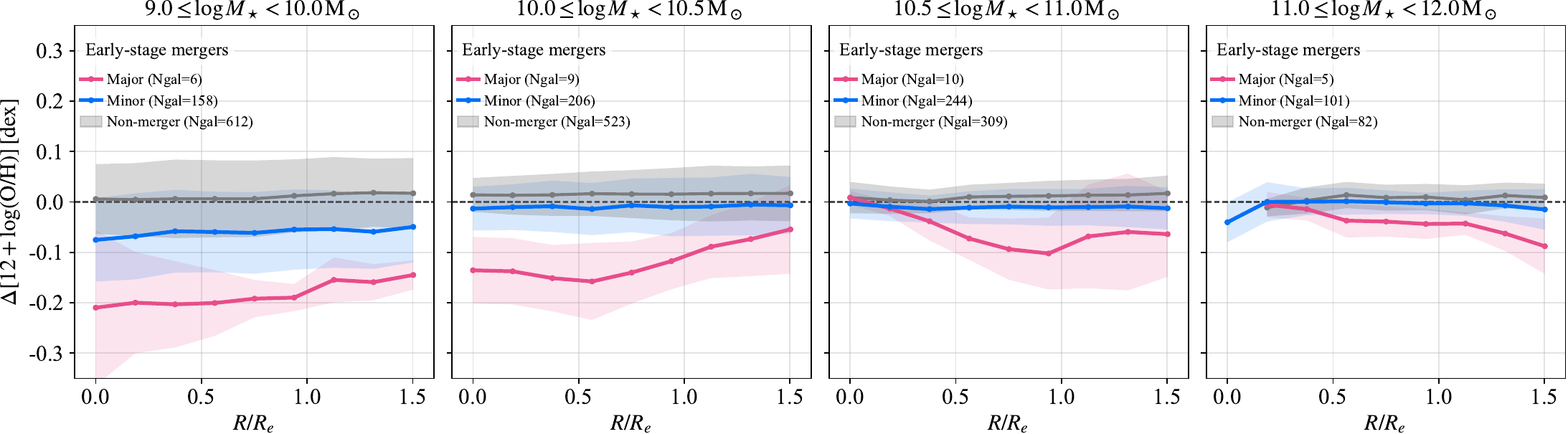}}
	\hspace{8pt}
	\subfigure[]{\label{fig_rad_oh_late}\includegraphics[width=0.9\textwidth]{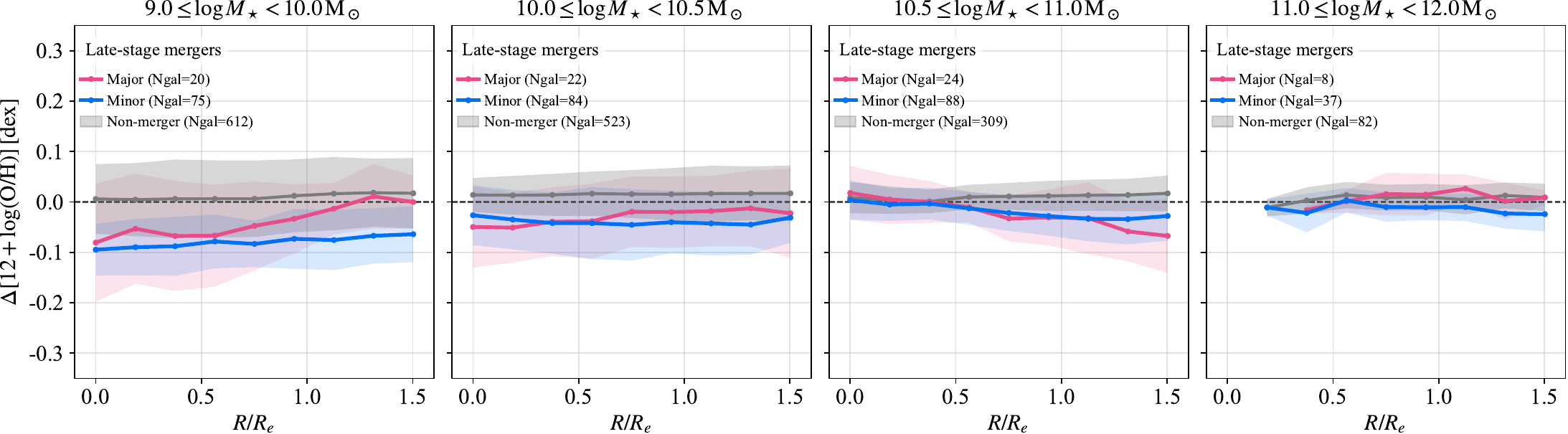}}
	\subfigure[]{\label{fig_rad_oh_post}\includegraphics[width=0.9\textwidth]{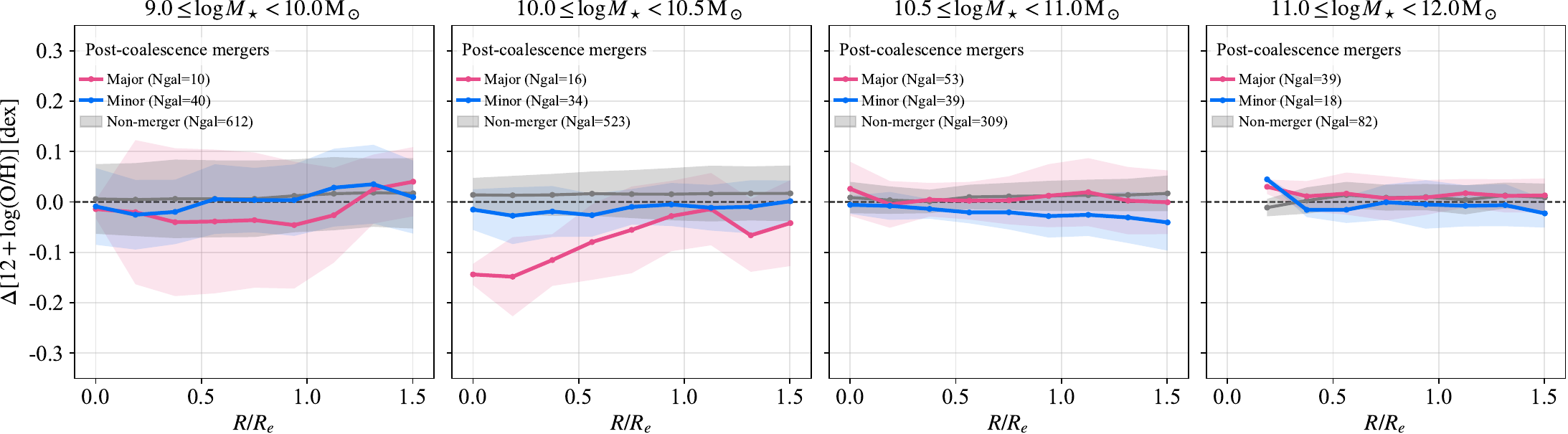}}
	\caption{Median radial profiles of resolved gas-phase metallicity offsets, \Doh, as a function of galactocentric radius normalized by the effective radius, $R/$\Reff. The plotting style and panel layout are the same as in Figure~\ref{fig_rad_ssfr}. Major mergers are shown in red, minor mergers in blue, and the non-merger control sample in gray. The shaded regions indicate the median absolute deviation (MAD) among galaxies in each radial bin, characterizing the galaxy-to-galaxy dispersion rather than the uncertainty in the median profile. Major mergers generally exhibit more negative \Doh\ than minor mergers, particularly in the early-stage, indicating stronger merger-driven metallicity dilution. A  stellar-mass dependence is seen in both major and minor mergers, with lower-\Mstar\ systems showing systematically stronger dilution than higher-\Mstar\ systems. Moreover, metallicity dilution is generally weaker in the post-coalescence phase than in the early and late stages for both merger types across most stellar-mass bins.
	}
	\label{fig_rad_oh}
\end{figure*}

Figure \ref{fig_rad_oh} presents the median radial profiles of \Doh\ as a function of galactocentric radius. The profiles are constructed using the same procedure described for \Dsf\ in Section \ref{sec_sfr_spx}. The shaded regions likewise show the MAD across galaxies in each radial bin and characterize the galaxy-to-galaxy dispersion about the median profiles.

For early-stage mergers (Figure \ref{fig_rad_oh_early}), major mergers show systematically negative \Doh\ across nearly all radial bins, indicating widespread gas-phase metallicity dilution relative to the control sample. This dilution is strongest in stellar mass bins with $\log M_\star/M_\odot < 11.0$, where the offsets reach $\sim -0.1$ to $-0.2$ dex. At the highest \Mstar, the metallicity offsets become weaker and approach zero. Minor mergers show a similar trend but with weaker dilution, typically less than $0.1$ dex. In both major and minor mergers, the strongest metallicity dilution is found in the lower-mass systems, with the effect weakening toward higher stellar masses. This \Mstar\ dependence is also reflected in the spatial extent of the dilution. In major mergers, the two lower-mass bins show negative \Doh\ extending across most of the radial range, whereas the dilution becomes weaker and less spatially extended at higher \Mstar. The lowest-mass minor mergers also show spatially extended dilution, although with smaller negative \Doh\ than the corresponding major mergers.

For late-stage mergers (Figure \ref{fig_rad_oh_late}), the overall radial trends, in terms of both profile shape and magnitude, for minor mergers are broadly similar to those in the early stage. A clear dependence on \Mstar\ remains, with the strongest dilution occurring in the lowest \Mstar\ bin ($\sim$ 0.05 -- 0.1 dex) and progressively weaker dilution toward higher masses ($\lesssim$ 0.05 dex). A similar mass-dependent trend is also present in major mergers. However, compared to the early-stage systems, the magnitude of metallicity dilution in major mergers is generally weaker than that observed in the early stage. For example, in the two lower-\Mstar\ bins, the central median \Doh\ is $\sim$ $-0.1$ dex for late-stage major mergers, compared with $\sim$ $-0.2$ dex for their early-stage counterparts. Consequently, the difference in \Doh\ between major and minor mergers is also reduced in the late stage, particularly in the two lower-\Mstar\ bins.

In post-coalescence systems (Figure \ref{fig_rad_oh_post}), the metallicity offsets are generally close to zero for both merger types, although appreciable dilution remains in some populations. The most prominent remaining dilution is found in major mergers at $10.0\leq\log M_\star/M_\odot<10.5$ (also evident in the \Doh\ spaxel distribution in Figure \ref{fig_dist_oh_post}), where \Doh\ reaches $\sim-0.15$ dex in the central region and becomes progressively less negative toward larger radii. In the other \Mstar\ bins, major mergers show only small offsets across most of the radial range, while minor mergers likewise exhibit little metallicity dilution at this stage.

We similarly quantify the radial trends in \Doh\ using Kendall's rank correlation test on the original unbinned spaxel measurements, with the results shown in Figure \ref{fig_rad_oh_stat_tau}. Many of the merger samples show statistically significant correlations according to our adopted criterion of $p<0.003$, although the correlation strengths are generally weak, with $|\tau|\lesssim0.22$. For early-stage major mergers, the two lower-mass bins show positive correlations ($\tau=0.117$ and $0.139$, both $p\leq0.001$), whereas the two higher-mass bins show negative correlations ($\tau=-0.101$ and $-0.210$, both $p\leq0.001$). Late-stage major mergers similarly show a positive correlation in the lowest-mass bin ($\tau=0.145$, $p\leq0.001$) and a negative correlation at $10.5\leq\log M_\star/M_\odot<11.0$ ($\tau=-0.166$, $p\leq0.001$), while the other two mass bins show little monotonic radial dependence. Following coalescence, the strongest correlation for major mergers occurs at $10.0\leq\log M_\star/M_\odot<10.5$, where $\tau=0.215$ ($p\leq0.001$), consistent with the increasingly less negative \Doh\ toward larger radii seen in the median profile. Minor mergers generally show weaker monotonic radial correlations, with $|\tau|\lesssim0.10$ across all merger stages and \Mstar\ bins. The direction of these correlations varies among the populations, with both positive and negative $\tau$ values and no common radial trend across merger stages or stellar masses. Overall, the Kendall tests show that the direction of the radial dependence of \Doh\ varies with merger type, merger stage, and \Mstar. Although many of these correlations are statistically significant, the generally small $|\tau|$ values indicate weak monotonic associations with radius.

In summary, the radial \Doh\ profiles show that metallicity dilution depends on \Mstar\ and merger stage. Both major and minor mergers generally show stronger dilution in the early stage than at later stages, with the effect being more pronounced in major mergers. For both merger types, the dilution is also generally stronger at lower \Mstar\ and weakens toward higher stellar masses. However, these trends are not strictly monotonic across all mass bins and merger stages, and some post-coalescence systems retain appreciable metallicity dilution. The Kendall tests identify statistically significant but generally weak radial correlations in many populations, with the direction of these correlations varying with merger type, merger stage, and \Mstar. Thus, there is no single radial dependence of \Doh\ common to all merger populations.

\section{Discussion}
\label{sec_discuss}
%The hierarchy is:
%* Early-stage merger: before second passage, typically after the galaxies become strongly interacting but before final approach. This includes the first-passage phase and the subsequent wider-separation phase.
%* Late-stage merger: between second passage and coalescence, when the nuclei are close and the system is strongly disturbed.
%* Post-coalescence merger: after the nuclei have merged into a single remnant.

\subsection{Merger-driven Evolution of Star Formation and Metallicity}

%verall framing → merger type → stage dependence of SFR → stage dependence of metallicity → mass dependence of metallicity → mass dependence of SFR → synthesis

\begin{figure*}
	\centering
	\includegraphics[width=0.95\textwidth]{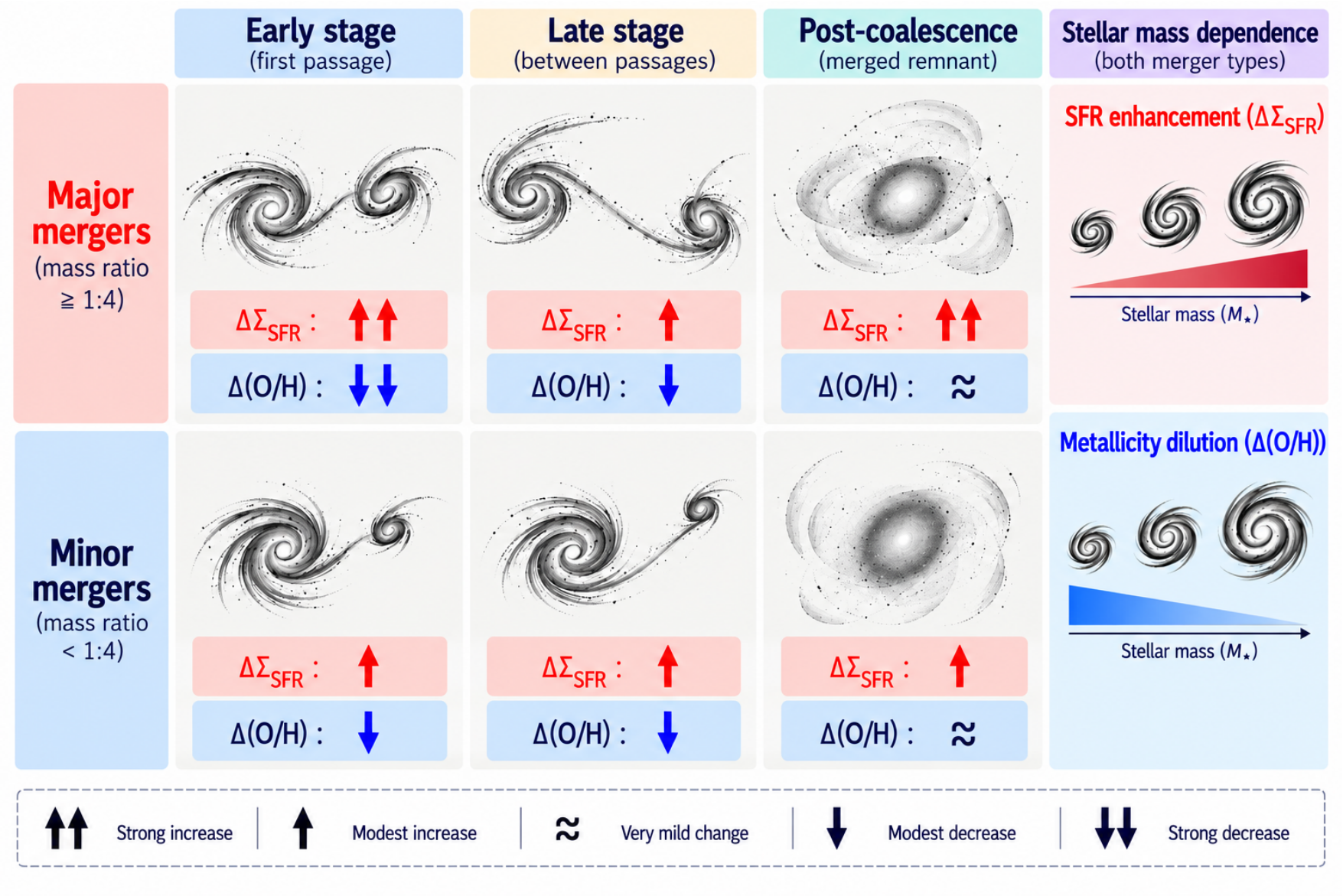}
	\caption{Schematic summary of the main trends in the resolved star formation and gas-phase metallicity responses to galaxy mergers. The middle three columns illustrate the qualitative changes in \Dsf\ and \Doh\ for major and minor mergers across the early, late, and post-coalescence stages, while the right column summarizes their general dependence on stellar mass. The arrow symbols indicate the qualitative strength and direction of the changes, as defined in the bottom legend, and are not intended to represent the quantitative magnitudes of the measured offsets. The galaxy illustrations are schematic and are not intended to represent the morphological characteristics of the galaxies in our sample.}
	\label{fig_summary}
\end{figure*}

Our results show that the resolved star formation and gas-phase metallicity offsets vary with merger type, merger stage, and stellar mass, as seen in both the distribution-based analyses (Figures~\ref{fig_dist_ssfr} and \ref{fig_dist_oh}) and the radial-profile analyses (Figures~\ref{fig_rad_ssfr} and \ref{fig_rad_oh}). The observed changes in \Dsf\ and \Doh\ are consistent with the expected effects of merger-driven gas redistribution, although our analysis does not directly trace gas flows or their temporal evolution. We therefore focus below on the observed differences among the merger populations and discuss possible physical interpretations in the context of previous studies.
Figure \ref{fig_summary} provides a schematic summary of the main trends identified in this work, illustrating the dependence of the resolved star formation and metallicity responses on merger type, merger stage, and stellar mass. The schematic is intended to represent the qualitative trends discussed in Section \ref{sec_result} rather than the quantitative magnitudes of the measured offsets. The galaxy illustrations are included for visualization purposes only and should not be interpreted as representing the morphological characteristics of the galaxies in our sample.

The resolved star formation and metallicity responses show a clear dependence on merger type. 
Major mergers generally exhibit larger positive \Dsf\ and more negative \Doh\ than minor mergers, although the magnitude of these differences varies with merger stage and stellar mass. The radial profiles further show that positive \Dsf\ in major mergers can extend over a substantial fraction of the galaxy, particularly in the early-stage and post-coalescence populations, whereas the enhancements in minor mergers are generally weaker. These results are broadly consistent with theoretical and simulation studies in which stronger tidal perturbations in major mergers produce more substantial gas redistribution \citep[e.g.,][]{Mih96,Cox08,Hop08,Tor12,Sch25}. 
Spatially extended star formation has also been observed in nearby interacting and merging systems, including the major-merger system of the Antennae galaxies \citep{Vig96,Mir98}, Arp 140 \citep{Cul06}, the NGC 2207/IC 2163 pair \citep{Elm06}, and nearby LIRG mergers \citep{Cor17}, which are typically associated with major mergers. 
However, while these observations provide a plausible connection between the spatially extended star formation enhancement found in our analysis and interaction-driven gas redistribution, our measurements do not directly constrain the strength or spatial extent of the underlying gas inflows.

The star formation response also varies with merger stage, particularly for major mergers. The \Dsf\ distributions and radial profiles show stronger enhancement in the early-stage and post-coalescence populations, whereas the late-stage population generally exhibits weaker enhancement. Thus, the observed star formation response does not follow a simple monotonic trend along the merger sequence. This behavior is qualitatively consistent with merger simulations that predict multiple episodes of enhanced star formation during an interaction \citep[e.g.,][]{Tor12,Bus18}. However, since the merger stages consist of different galaxy populations rather than observations of the same systems at successive times, we cannot directly trace the temporal evolution of individual star formation episodes. Nevertheless, the weaker enhancement in the late-stage population followed by stronger enhancement in the post-coalescence population is broadly consistent with an inter-burst phase followed by renewed merger-induced star formation.

The metallicity response shows a different dependence on merger stage. The strongest negative \Doh\ values are generally found in early-stage major mergers, particularly at lower stellar masses, while the offsets tend to decrease at later stages and at higher stellar masses. However, this evolution is not strictly monotonic, and some post-coalescence populations retain appreciable metallicity dilution. The radial profiles show relatively modest variations in \Doh\ with radius, while the Kendall tests indicate generally weak correlations whose direction varies among merger stages and stellar-mass bins. Thus, there is no common radial dependence of \Doh\ across the merger populations.

These results are consistent with a scenario in which interaction-driven redistribution of lower-metallicity gas contributes to the observed dilution, while subsequent star formation, chemical enrichment, and gas mixing further modify the metallicity distribution as the merger progresses. These processes may also account for the less systematic radial structure of \Doh\ compared with \Dsf. Gas compression and accumulation can produce relatively coherent spatial variations in star formation, including stronger central enhancement and, in some early-stage and post-coalescence major mergers, spatially extended enhancement. In contrast, \Doh\ reflects the combined effects of gas redistribution, the pre-existing metallicity distribution, radial mixing, and chemical enrichment. Variations in the relative importance of these processes among merger populations may contribute to the more diverse radial behavior of \Doh. Our measurements, however, do not distinguish among these processes or establish their relative contributions.

A clear stellar-mass dependence is particularly apparent in the metallicity response. Lower-mass systems generally show larger negative \Doh\ than higher-mass systems in both major and minor mergers, although the magnitude of this dependence varies with merger stage. Several galaxy properties could contribute to this behavior. Previous studies have shown systematic variations in metallicity gradients \citep{Bel17,San20} and gas fractions \citep{Pee14,Sai17,Lap20} with galaxy properties, both of which can affect the metallicity response to radial gas redistribution. 
Differences in gravitational potential may also play a role. Lower-mass galaxies generally have shallower gravitational potentials, which could make their gas more susceptible to perturbation and redistribution during an interaction. Such redistribution could mix lower-metallicity gas from larger radii with the more metal-rich gas in the inner regions, producing a larger negative \Doh. Conversely, the deeper gravitational potentials of more massive galaxies could reduce the response of their gas distributions to comparable tidal perturbations.

The star formation response exhibits a different stellar-mass dependence. The largest positive \Dsf\ values are generally found in intermediate- and high-mass major mergers, particularly in the early-stage and post-coalescence populations, whereas the lowest-mass major mergers show comparatively weaker enhancement. This contrasts with the stronger metallicity dilution observed toward lower stellar masses and indicates that the amplitudes of the star formation and metallicity responses do not scale with stellar mass in the same way. 
Simulations have suggested that the gas fraction can influence the efficiency of merger-induced star formation \citep{Cox08,Scu15,Fen17}. Gas fraction generally decreases with increasing stellar mass \citep[e.g.,][]{Sai17}. In gas-rich galaxies, strong turbulence, fragmentation, and star formation may already be present prior to an interaction, such that a merger produces a smaller additional enhancement in star formation. In contrast, in systems with lower gas fractions, turbulence, fragmentation, and gas inflows can be more strongly enhanced by the interaction, resulting in a larger relative increase in star formation \citep{Scu15,Fen17}. This dependence on gas fraction could therefore contribute to the comparatively weaker \Dsf\ enhancement observed in our lower-mass systems, although we do not directly measure gas fractions in the present analysis.

Overall, our results show that the resolved star formation and metallicity responses to galaxy mergers depend jointly on merger type, merger stage, stellar mass, and galactocentric radius. The different behaviors of \Dsf\ and \Doh\ show that the star formation and metallicity responses have different dependencies on merger stage and stellar mass. These trends are broadly consistent with an interplay among gas redistribution, star formation, and chemical evolution, although distinguishing the relative roles and timescales of these processes will require direct constraints on the gas distribution and kinematics.

\subsection{Caveats}

Several caveats should be noted in this analysis.
First, our resolved analysis is restricted to star-forming spaxels, since star formation and \OH\ measurements can only be reliably derived  in regions with ongoing star formation. In principle, $\Sigma_{\mathrm{SFR}}$ could also be estimated for non-star-forming spaxels using the D4000 calibration \citep{Blu20}. However, these estimates are known to have substantially larger uncertainties \citep[see Figure 4 in][]{Blu20}. Since our sample is already restricted to globally star-forming galaxies, the fraction of non-star-forming spaxels is relatively small, so excluding them is unlikely to qualitatively affect our main conclusions.

Second, by selecting only globally star-forming galaxies, our sample is biased against bulge-dominated and early-type systems, which are more likely to be quiescent. This means that our results primarily reflect the impact of mergers on actively star-forming disk galaxies. Previous studies have shown that mergers can also affect star formation in early-type systems \citep[e.g.,][]{Dav13,Dav15,Hwa15,Van18}, for example through changes in ISM properties driven by merger-induced gas redistribution or through dynamical effects that stabilize the gas against gravitational collapse. Such effects are not fully captured in our current analysis and should be explored in future work with a broader galaxy sample.

Third, our analysis does not explicitly identify the companion galaxy or include information on its properties, such as whether it is star-forming or quiescent.  Consequently, we do not distinguish between different merger configurations, such as wet mergers between two star-forming galaxies and mixed mergers between a star-forming and a quiescent galaxy. This may introduce additional diversity in the merger-driven response, as interactions with gas-rich, star-forming companions could trigger stronger gas inflows and greater star formation enhancement than interactions with gas-poor or quiescent companions \citep{Lin07,He22,Bro23,Cal24,Shi26}. In mixed mergers, the hot circumgalactic medium (CGM) associated with an early-type companion may also suppress the accretion of cold gas onto the star-forming galaxy, further reducing its star formation activity \citep{Hwa15}. Such differences may also partly reflect changes in gas properties under different merger configurations (e.g., wet versus dry mergers), which can further regulate the efficiency of star formation \citep{Cao16,Lis19}.

Finally, our merger classifications rely on the MaNGA Galaxy Merger Catalog \citep{Com24}, which is based on the supervised machine-learning framework of \citet{Nev23}. Although this catalog provides a uniform and physically motivated classification of merger type and stage, it is ultimately limited by the assumptions of the underlying simulations, the realism of the mock observations, and the ability of morphological indicators to distinguish mergers from non-mergers. In particular, merger signatures may be missed in systems with weak tidal features, unfavorable viewing angles, or advanced dynamical relaxation, while some non-merging galaxies with irregular or asymmetric morphologies may be incorrectly identified as mergers. The assigned probabilities therefore represent statistical classifications rather than definitive merger identifications, and these uncertainties should be kept in mind when interpreting the results.

\section{Summary}
\label{sec_summary}
In this work, we investigate how galaxy mergers affect resolved star formation and gas-phase metallicity, with a particular focus on merger type (major versus minor), merger stage, and the stellar mass of the host galaxies. We use the spatially resolved MaNGA data products to measure star formation and gas-phase metallicity, together with the MaNGA Galaxy Merger Catalog developed by \citet{Nev23} and \citet{Com24} to classify galaxies by merger type and evolutionary stage.
By constructing matched control samples at both the galaxy and spaxel levels, we quantify local offsets in star formation rate surface density (\Dsf) and gas-phase metallicity (\Doh) relative to non-merger analogs. This allows us to isolate merger-driven perturbations while controlling for both global galaxy properties and local resolved environments. Our main results can be summarized as follows:

\begin{enumerate}
	\item \emph{Dependence on merger type.} The magnitude of merger-driven perturbations depends on merger type. Major mergers generally show stronger star formation enhancement and metallicity dilution than minor mergers, although the magnitude of these differences varies with merger stage and stellar mass. Both the distribution-based (Section \ref{sec_result_spx_dist} and Figures~\ref{fig_dist_ssfr} and \ref{fig_dist_oh}) and radial-profile analyses (Section \ref{sec_result_radial} and Figures~\ref{fig_rad_ssfr} and \ref{fig_rad_oh}) show larger positive \Dsf\ and more negative \Doh\ in many major-merger populations than in their minor-merger counterparts. These differences are consistent with a stronger resolved response of star formation and metallicity to major mergers.
	
	\item \emph{Dependence of star formation on merger stage.} The merger-driven star formation response depends on merger stage. Major mergers show a non-monotonic dependence of star formation enhancement on merger stage. The early-stage population generally shows enhanced \Dsf, while the late-stage population shows weaker enhancement. Strong enhancement is again observed in post-coalescence major mergers, particularly at intermediate and high stellar masses. The median \Dsf\ values of early-stage major mergers range from $\sim$ $0.13$ to $0.23$ dex across the four \Mstar\ bins, while post-coalescence major mergers reach $\sim$ $0.26$ and $0.40$ dex in the $10.0$--$10.5$ and $11.0$--$12.0$ \Mstar\ bins, respectively. In contrast, minor mergers generally show median \Dsf\ values below $0.13$ dex and exhibit less variation with merger stage than major mergers. Thus, the star formation response does not follow a simple monotonic trend across the merger sequence (Section \ref{sec_sfr_spx} and Figure \ref{fig_dist_ssfr}).
	
	\item \emph{Dependence of metallicity on merger stage.} The merger-driven metallicity response also depends on merger stage. The strongest metallicity dilution is found in early-stage major mergers at lower stellar masses, with median \Doh\ values of $-0.15$ and $-0.14$ dex in the two lowest-\Mstar\ bins, compared with $\sim-0.05$ and $-0.04$ dex for minor mergers. In late-stage and post-coalescence systems, the median offsets generally remain negative, but the metallicity dilution is less than $\sim0.1$ dex for both merger types in most \Mstar\ bins. Overall, metallicity dilution tends to weaken toward the post-coalescence stage, although the stage dependence differs between major and minor mergers and is not monotonic across all \Mstar\ bins (Section \ref{sec_oh_spx} and Figure \ref{fig_dist_oh}).
	
	\item \emph{Radial dependence.}  Merger-driven perturbations show different spatial behaviors in star formation and metallicity.
	Star formation enhancement in major mergers can extend over a substantial fraction of the galaxy, particularly in the early-stage and post-coalescence populations, whereas the enhancement in minor mergers is generally weaker and more centrally concentrated. In contrast, the radial behavior of metallicity is more complex, with the strength and direction of the radial trends varying with merger stage and stellar mass and showing no common radial dependence across the merger populations (Section \ref{sec_result_radial} and Figures \ref{fig_rad_ssfr} and \ref{fig_rad_oh}).
	
	\item \emph{Dependence on stellar mass.} The response of star formation and metallicity depends on stellar mass.
	Lower-mass galaxies generally exhibit stronger metallicity dilution in both major and minor mergers, although the strength of this dependence varies with merger stage. In contrast, the strongest star formation enhancement is generally found in intermediate- and high-mass major mergers, particularly in the early-stage and post-coalescence populations. Thus, the star formation and metallicity responses show different dependencies on stellar mass (Section \ref{sec_result}).
\end{enumerate}

In summary, our results show that the resolved star formation and metallicity responses to galaxy mergers depend jointly on merger type, merger stage, stellar mass, and galactocentric radius. Major mergers generally show stronger responses than minor mergers, but neither \Dsf\ nor \Doh\ follows a simple monotonic trend across the merger sequence, and the two quantities exhibit different dependencies on stellar mass. These observational trends are consistent with an interplay among merger-driven gas redistribution, star formation, and chemical evolution, although our measurements do not directly constrain the underlying gas flows or their timescales. Future work extending this analysis to quiescent and early-type galaxies and incorporating molecular gas measurements will provide further insight into how mergers regulate the baryon cycle and influence galaxy evolution across a broader range of galaxy properties, merger configurations, and ISM conditions.

%\bibliography{sample701}{}
%\bibliographystyle{aasjournalv7}
\begin{acknowledgments}
We thank the anonymous referee and the statistics editor for their careful reading and constructive feedback, which helped improve the quality and clarity of this manuscript.
HAP acknowledges support from the National Science and Technology Council of Taiwan under grant  113-2112-M-032-014-MY3 and thanks Lihwai Lin for preparing the MaNGA data products in a format that facilitated the analysis presented in this work.

This project makes use of the MaNGA-Pipe3D dataproducts. We thank the IA-UNAM MaNGA team for creating these dataproducts, and the Conacyt Project CB-285080 for supporting them.

Funding for the Sloan Digital Sky Survey IV has been provided by the 
Alfred P. Sloan Foundation, the U.S.  Department of Energy Office of 
Science, and the Participating  Institutions. 
SDSS-IV acknowledges support and  resources from the Center for High 
Performance Computing  at the  University of Utah. The SDSS website is www.sdss4.org.
SDSS-IV is managed by the Astrophysical Research Consortium for the Participating Institutions of the SDSS Collaboration including the Brazilian Participation Group, the Carnegie Institution for Science, Carnegie Mellon University, Center for Astrophysics | Harvard \& Smithsonian, the Chilean Participation Group, the French Participation Group, Instituto de Astrof\'isica de Canarias, The Johns Hopkins University, Kavli Institute for the Physics and Mathematics of the Universe (IPMU) / University of Tokyo, the Korean Participation Group, Lawrence Berkeley National Laboratory, Leibniz Institut f\"ur Astrophysik Potsdam (AIP), Max-Planck-Institut f\"ur Astronomie (MPIA Heidelberg), Max-Planck-Institut f\"ur Astrophysik (MPA Garching), Max-Planck-Institut f\"ur Extraterrestrische Physik (MPE), National Astronomical Observatories of China, New Mexico State University, New York University, University of Notre Dame, Observat\'ario Nacional / MCTI, The Ohio State University, Pennsylvania State University, Shanghai Astronomical Observatory, United Kingdom Participation Group, Universidad Nacional Aut\'onoma de M\'exico, University of Arizona, University of Colorado Boulder, University of Oxford, University of Portsmouth, University of Utah, University of Virginia, University of Washington, University of Wisconsin, Vanderbilt University, and Yale University.

\end{acknowledgments}

\appendix
\section{Distribution-shape statistics}
\label{appendix_distribution_stats}

\setcounter{figure}{0}
\renewcommand{\thefigure}{A\arabic{figure}}
\renewcommand{\theHfigure}{A\arabic{figure}}

\makeatletter
\renewcommand{\p@subfigure}{\thefigure}
\renewcommand{\thesubfigure}{(\alph{subfigure})}
\makeatother
%Skewness → asymmetry / which side has the tail
%Excess kurtosis → tail weight relative to a Gaussian
%D’Agostino–Pearson \(K^2\) → whether the distribution is consistent with normality
%Hartigan’s Dip test → whether there is evidence against a unimodal distribution

To complement the Anderson--Darling two-sample tests of the \Dsf\ and \Doh\ distributions presented in Section \ref{sec_result_spx_dist}, we further characterize their distribution shapes using skewness, excess kurtosis, the D'Agostino--Pearson \(K^2\) normality test, and the Hartigan Dip test for unimodality. Skewness measures the asymmetry of a distribution; positive values indicate a more extended tail toward positive offsets, while negative values indicate a more extended tail toward negative offsets. For a Gaussian distribution, skewness is zero. Absolute skewness values between 0.5 and 1 are generally considered to indicate moderate asymmetry, while values exceeding 1 indicate substantial asymmetry \citep{Bul79}. We report excess kurtosis using the Fisher definition, for which a Gaussian distribution has a value of zero. Positive excess kurtosis indicates heavier tails and a greater prevalence of extreme values relative to a Gaussian distribution, whereas negative excess kurtosis indicates lighter tails and fewer extreme values. 
Although no universally accepted thresholds exist for excess kurtosis, absolute values exceeding approximately 1 may indicate noticeable departures from Gaussian tail behavior.
These numerical guidelines provide a descriptive assessment of distribution shape, while the statistical significance of departures from Gaussianity is assessed using the D'Agostino--Pearson \(K^2\) test described below.

The D'Agostino--Pearson \(K^2\) test evaluates the null hypothesis that a sample is drawn from a Gaussian distribution by jointly considering departures in skewness and kurtosis. The Dip test provides a complementary assessment of distribution shape by testing the null hypothesis that the distribution is unimodal; a small \(p\)-value therefore indicates evidence for departure from unimodality. For both tests, we adopt the same significance criterion of \(p<0.003\) used for the distribution comparisons in the main text. Since the sensitivity of the \(K^2\) test increases with sample size, relatively modest departures from Gaussianity can produce large \(K^2\) statistics and small \(p\)-values in very large samples. The absolute \(K^2\) values should therefore not be interpreted as a direct measure of the magnitude of departures from normality when comparing populations with substantially different numbers of spaxels, as is the case for the merger and control samples shown in Figures~\ref{fig_dist_ssfr} and \ref{fig_dist_oh}. We instead use the \(K^2\) test to assess whether individual distributions are consistent with Gaussianity, while skewness and excess kurtosis characterize the form and magnitude of their departures from a Gaussian shape.

\subsection{Resolved star formation rate offsets}
\label{appendix_sf}

\begin{figure*}[!ht]
	\centering
	\subfigure[]{\label{fig_sf_stat_skewness}\includegraphics[width=0.45\textwidth]{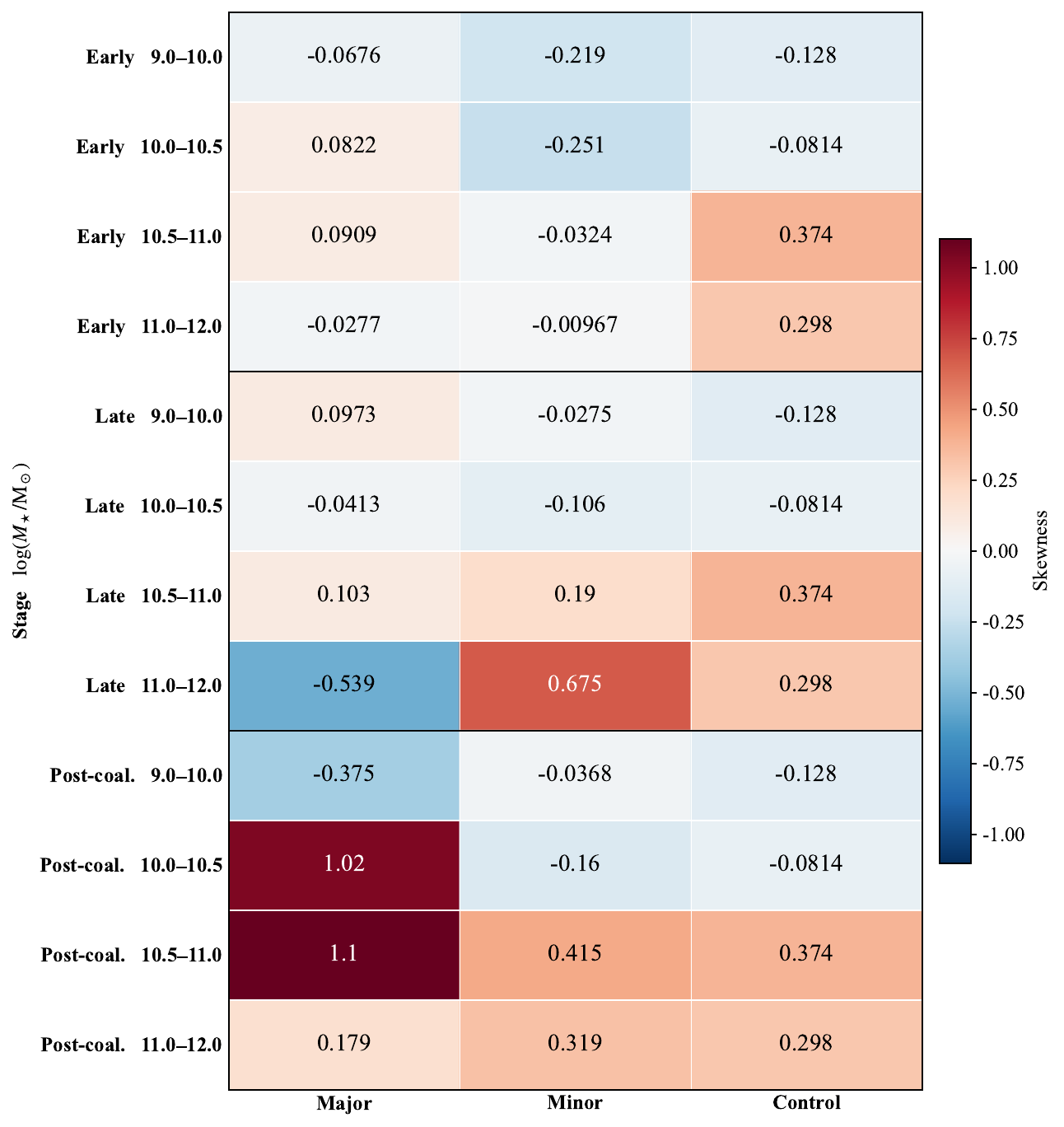}}
	\subfigure[]{\label{fig_sf_stat_kurtosis}\includegraphics[width=0.45\textwidth]{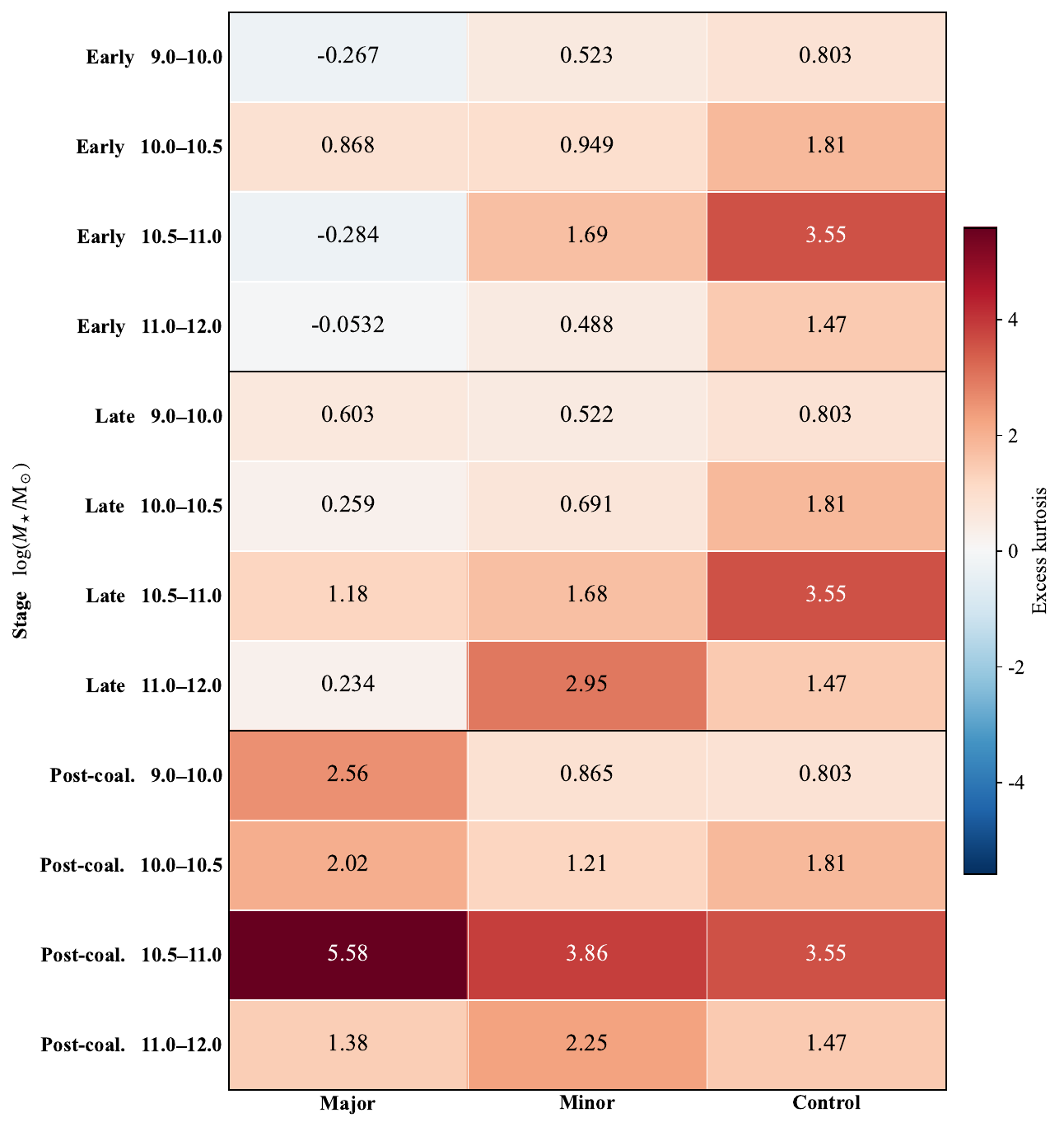}}
	\subfigure[]{\label{fig_sf_stat_k2}\includegraphics[width=0.45\textwidth]{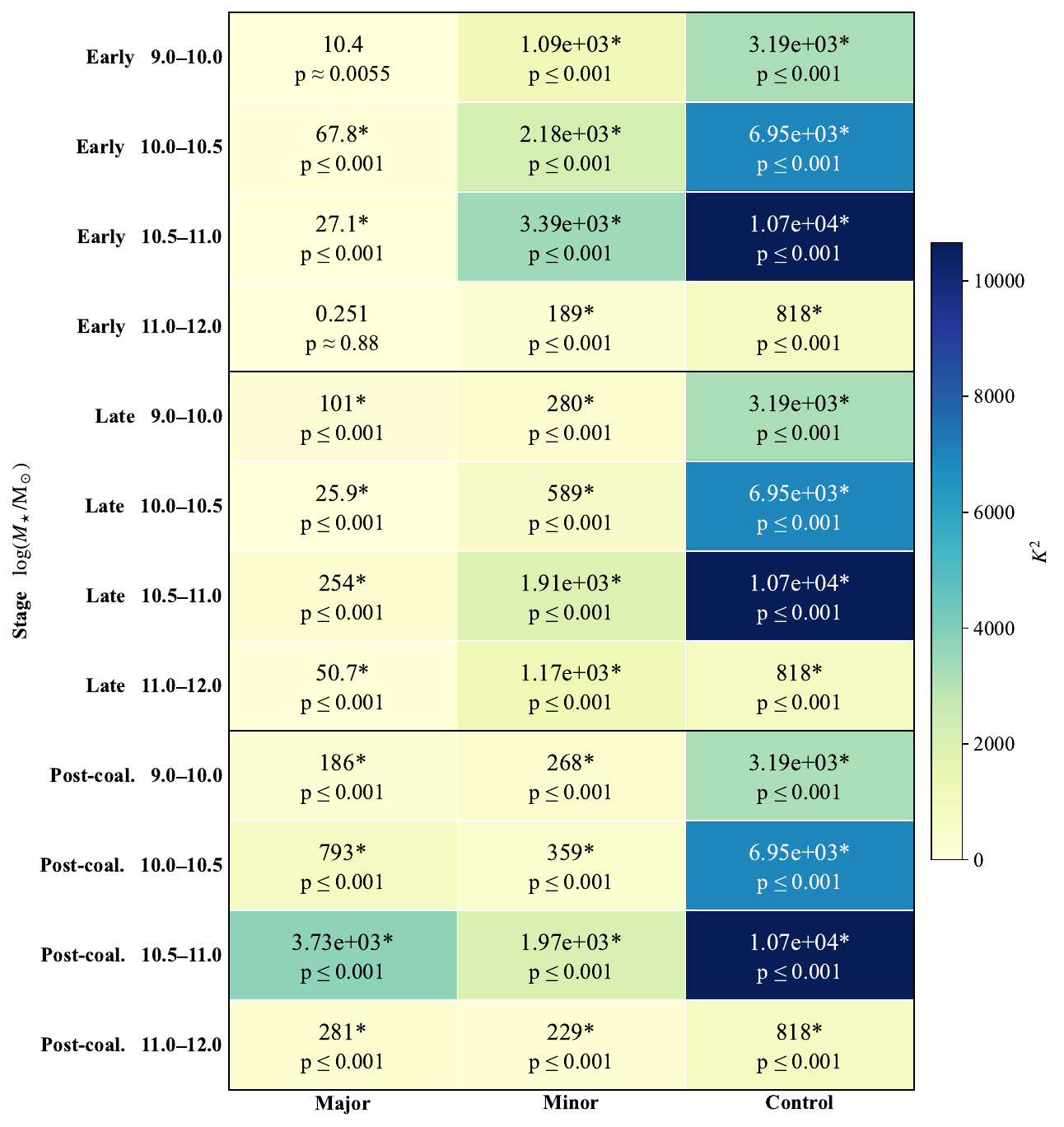}}
	\subfigure[]{\label{fig_sf_stat_dip}\includegraphics[width=0.45\textwidth]{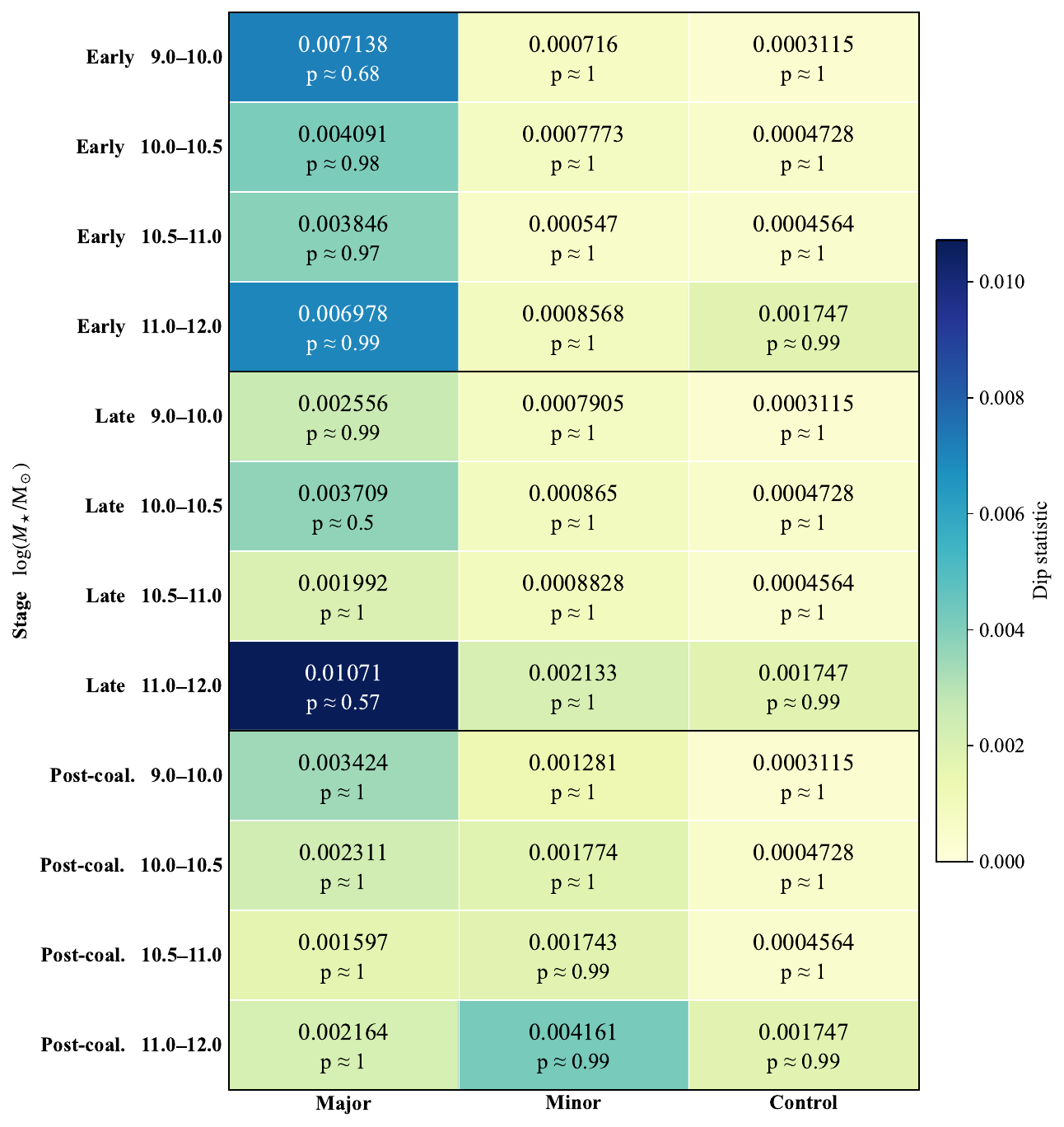}}
	\caption{Distribution-shape statistics for \Dsf\ of major mergers, minor mergers, and their matched control galaxies, shown separately for the early, late, and post-coalescence merger stages and in four stellar-mass bins. {\bf Panel (a):} Skewness and {\bf Panel (b):} excess kurtosis (Fisher definition) of each distribution. Positive (negative) skewness indicates a distribution with a more extended high-\Dsf\ (low-\Dsf) tail, while positive (negative) excess kurtosis indicates heavier (lighter) tails relative to a normal distribution. {\bf Panel (c):} \(K^2\) statistic from the D'Agostino--Pearson normality test. {\bf Panel (d):} Dip statistic from the Hartigan Dip test for unimodality. The values in panels (c) and (d) give the corresponding test statistic and associated \(p\)-value. Asterisks indicate distributions that reject the corresponding null hypothesis at the adopted significance criterion of \(p<0.003\).}
	\label{fig_sf_stat_all}
\end{figure*}

Figure \ref{fig_sf_stat_all} presents the distribution-shape statistics for \Dsf. The early- and late-stage major-merger distributions generally show only modest asymmetry, with skewness values close to zero in most bins (Figure \ref{fig_sf_stat_skewness}). The main exception is the highest-mass late-stage bin, which has a skewness of \(-0.54\).  

The most prominent positive asymmetry occurs after coalescence.
The major-merger distributions in the \(10.0\leq\log(M_\star/M_\odot)<10.5\) and \(10.5\leq\log(M_\star/M_\odot)<11.0\) bins have skewness values of \(1.02\) and \(1.10\), respectively, quantitatively supporting the extended positive-\Dsf\ tails visible in Figure \ref{fig_dist_ssfr}. The corresponding excess kurtosis values (Figure \ref{fig_sf_stat_kurtosis}) are \(2.02\) and \(5.58\), indicating substantial tail weight relative to a Gaussian distribution, particularly in the latter bin. In contrast, the post-coalescence major-merger distribution in the lowest-mass bin has negative skewness (\(-0.375\)), while the highest-mass bin shows only weak positive skewness (\(0.179\)).

The minor-merger distributions show a different pattern. Their skewness is generally modest in the early-stage and lower-mass late-stage bins, but reaches \(0.675\) in the highest-mass late-stage bin and \(0.415\) and \(0.319\) in the two highest-mass post-coalescence bins, respectively. The excess kurtosis is positive in all displayed bins, ranging from \(0.488\) to \(3.86\), with the largest value found in the \(10.5\leq\log(M_\star/M_\odot)<11.0\) post-coalescence population.

The \(K^2\) statistics and their associated \(p\)-values are shown in Figure \ref{fig_sf_stat_k2}. Using our adopted significance criterion of \(p<0.003\), most \Dsf\ distributions show statistically significant departures from normality. For major mergers, the exceptions are the lowest- and highest-\Mstar\ early-stage bins, with \(p\simeq0.0055\) and \(p\simeq0.88\), respectively. All other major-merger distributions, as well as all minor-merger and control distributions, satisfy the adopted criterion for rejecting normality.

The Dip statistics and their associated \(p\)-values are shown in Figure \ref{fig_sf_stat_dip}. None of the major-merger, minor-merger, or control \Dsf\ distributions reject unimodality at our adopted significance criterion of \(p<0.003\). 
Thus, although the \(K^2\) tests identify significant departures from Gaussianity in most distributions, the Dip tests provide no significant evidence for multimodality. The departures from Gaussianity are instead accompanied by varying degrees of asymmetry and tail weight, as quantified by the skewness and excess kurtosis.

The matched control distributions also show non-zero skewness and excess kurtosis, demonstrating that asymmetric and heavy-tailed offset distributions are not unique to the merger populations. The control distributions therefore provide an empirical reference for assessing merger-related changes in distribution shape. However, the numbers of spaxels differ substantially among the major-merger, minor-merger, and control samples (Figure \ref{fig_dist_ssfr}). These differences in sample size should be considered when interpreting the statistical significance of the distribution-shape statistics, since the sensitivity of statistical tests generally increases with sample size.

Overall, these distribution-shape statistics support the trends identified from the \Dsf\ distributions in Section \ref{sec_sfr_spx} and Figure \ref{fig_dist_ssfr}. The most prominent merger-related differences in distribution shape are found for post-coalescence major mergers, where the large positive skewness and excess kurtosis quantify the extended high-\Dsf\ tails. Minor mergers generally show more modest asymmetries. Although departures from Gaussianity are common across both merger and control populations, the Dip tests provide no significant evidence for multimodality. Thus, the \Dsf\ distributions are characterized by varying degrees of asymmetry and tail weight rather than the emergence of distinct multiple populations.

\subsection{Resolved gas-phase metallicity offsets}
\label{appendix_oh}
\begin{figure*}[!ht]
	\centering
	\subfigure[]{\label{fig_oh_stat_skewness}\includegraphics[width=0.45\textwidth]{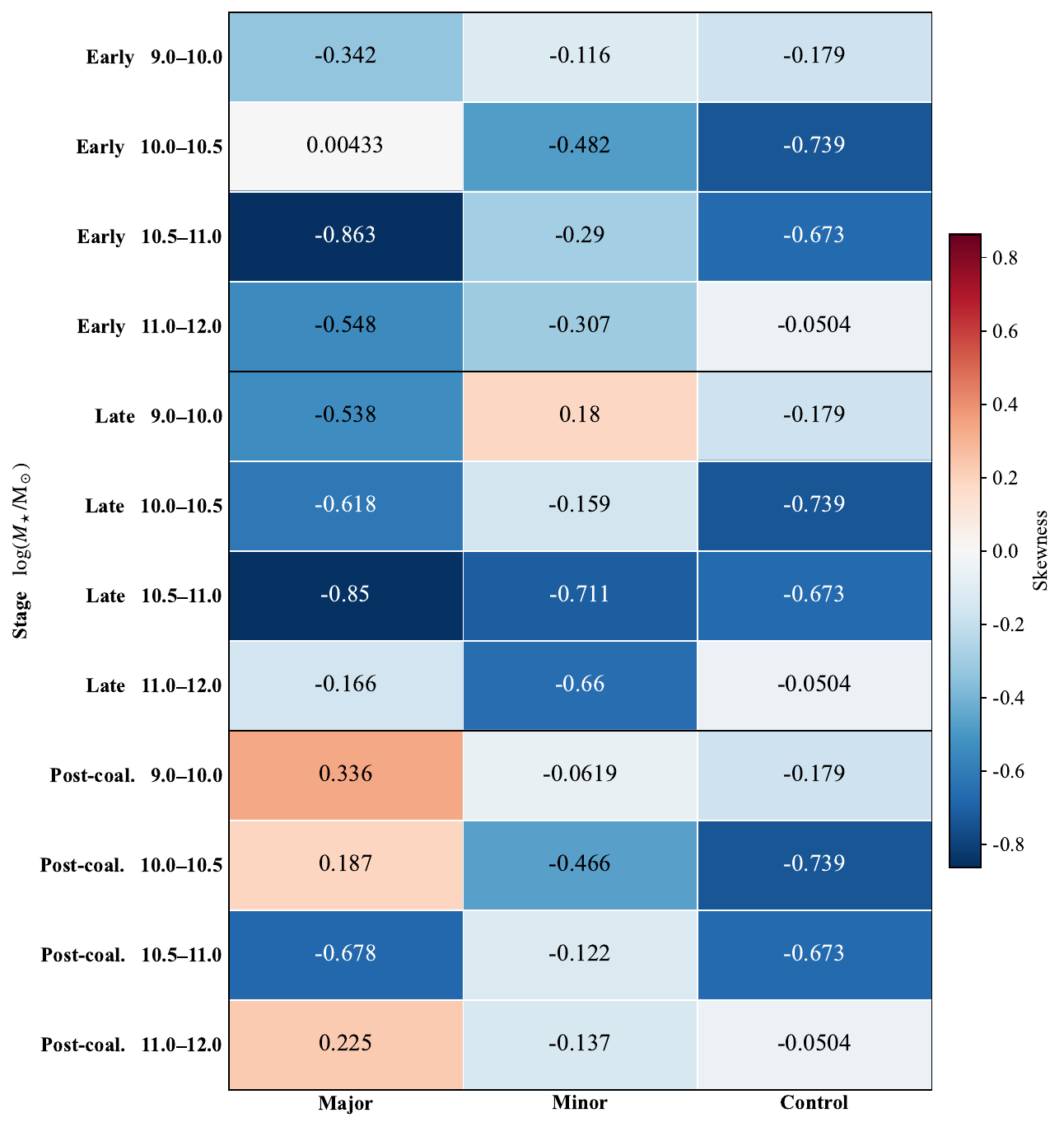}}
	\subfigure[]{\label{fig_oh_stat_kurtosis}\includegraphics[width=0.45\textwidth]{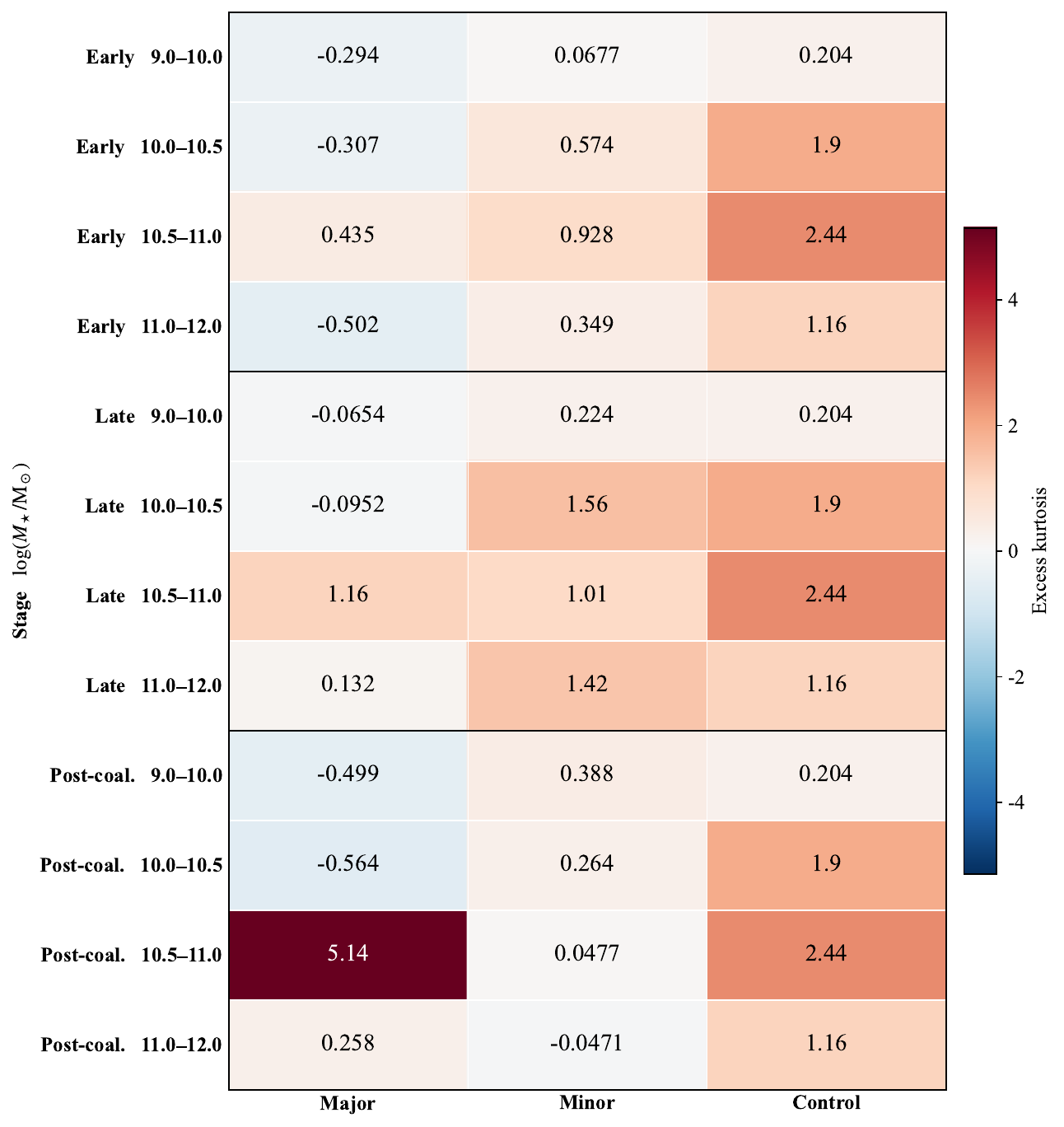}}
	\subfigure[]{\label{fig_oh_stat_k2}\includegraphics[width=0.45\textwidth]{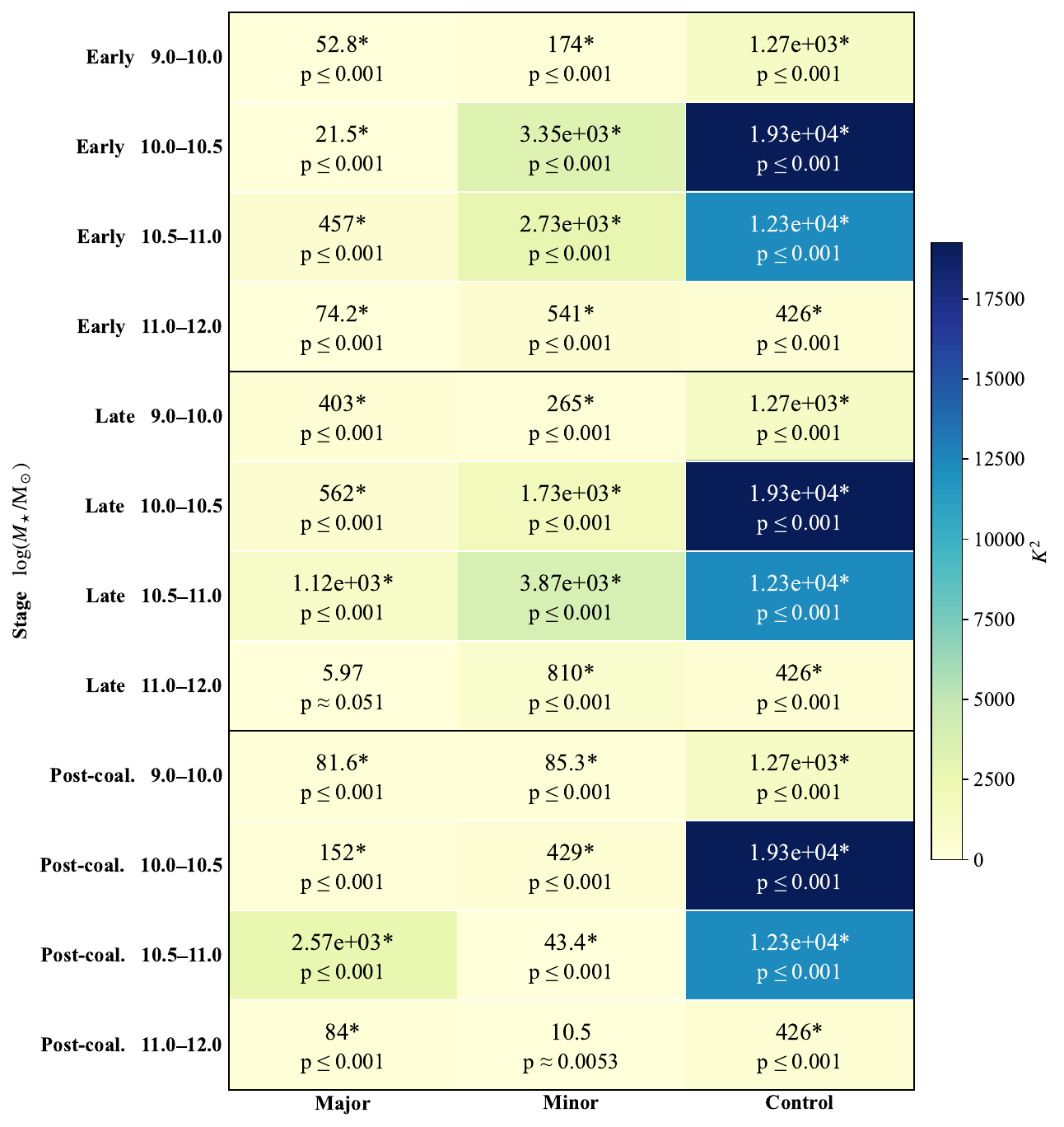}}
	\subfigure[]{\label{fig_oh_stat_dip}\includegraphics[width=0.45\textwidth]{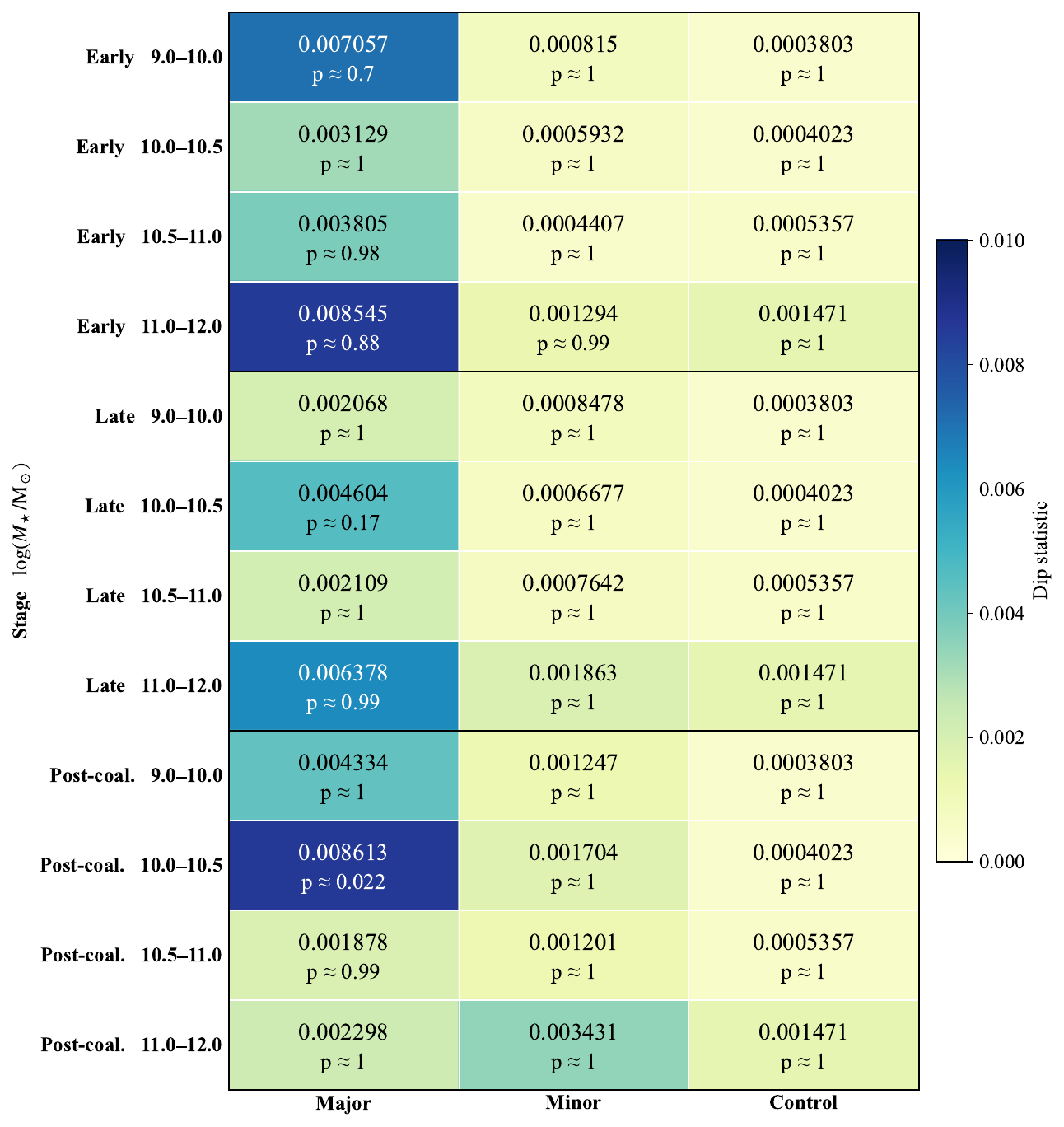}}
	\caption{ Distribution-shape statistics for the metallicity offset (\Doh) of major mergers, minor mergers, and their matched control galaxies, shown separately for the early, late, and post-coalescence merger stages and in four stellar-mass bins. {\bf Panel (a):} Skewness and {\bf Panel (b):} excess kurtosis (Fisher definition) of each distribution. Positive (negative) skewness indicates a distribution with a more extended high-\Doh\ (low-\Doh) tail, while positive (negative) excess kurtosis indicates heavier (lighter) tails relative to a normal distribution. {\bf Panel (c):} \(K^2\) statistic from the D'Agostino--Pearson normality test. {\bf Panel (d):} Dip statistic from the Hartigan Dip test for unimodality. The values in panels (c) and (d) give the corresponding test statistic and associated \(p\)-value. Asterisks indicate distributions that reject the corresponding null hypothesis at the adopted significance criterion of \(p<0.003\).}
	\label{fig_oh_stat_all}
\end{figure*}

Figure \ref{fig_oh_stat_all} presents the corresponding distribution-shape statistics for \Doh. In contrast to \Dsf, the metallicity-offset distributions of major mergers are predominantly negatively skewed during the early and late merger stages (Figure \ref{fig_oh_stat_skewness}). Early-stage major mergers have skewness values of \(-0.342\), \(0.004\), \(-0.863\), and \(-0.548\) with increasing stellar mass, while the corresponding late-stage values are \(-0.538\), \(-0.618\), \(-0.850\), and \(-0.166\). The negative skewness in most of these bins indicates extended tails toward low \Doh, consistent with the metallicity-diluted tails visible in Figure \ref{fig_dist_oh}. Minor mergers also frequently show negative skewness, although its magnitude and dependence on stellar mass vary with merger stage.

The distribution shapes change after coalescence. Major mergers show positive skewness in the two lowest-mass bins (\(0.336\) and \(0.187\)), negative skewness in the \(10.5\leq\log(M_\star/M_\odot)<11.0\) bin (\(-0.678\)), and weak positive skewness in the highest-mass bin (\(0.225\)). The \(10.5\leq\log(M_\star/M_\odot)<11.0\) post-coalescence major-merger population is particularly distinctive, with an excess kurtosis of \(5.14\) (Figure \ref{fig_oh_stat_kurtosis}), indicating substantial tail weight relative to a Gaussian distribution. The minor-merger distributions generally show  modest excess kurtosis, ranging from $\sim$ \(-0.05\) to \(1.56\), and are predominantly negatively skewed, with the lowest-mass late-stage population being the main exception.

The \(K^2\) statistics and their associated \(p\)-values are shown in Figure \ref{fig_oh_stat_k2}. Most \Doh\ distributions depart significantly from normality. For major mergers, all distributions satisfy our adopted \(p<0.003\) criterion except the highest-mass late-stage population (\(p\simeq0.051\)). For minor mergers, all distributions satisfy the criterion except the highest-mass post-coalescence population (\(p\simeq0.0053\)). All matched control distributions also show statistically significant departures from normality (\(p\leq0.001\)).

The Dip statistics and their associated \(p\)-values are shown in Figure \ref{fig_oh_stat_dip}. None of the major-merger, minor-merger, or control \Doh\ distributions reject unimodality at our adopted significance criterion of \(p<0.003\). The smallest \(p\)-value occurs for the \(10.0\leq\log(M_\star/M_\odot)<10.5\) post-coalescence major-merger population (\(p\simeq0.022\)), which remains above the adopted threshold. Thus, although most \Doh\ distributions show statistically significant departures from Gaussianity, there is no significant evidence for multimodality.

As for \Dsf, the control \Doh\ distributions also show non-zero skewness and excess kurtosis, demonstrating that departures from a Gaussian shape are not unique to the merger populations. Again, given the different numbers of spaxels among the populations (Figure \ref{fig_dist_oh}), the distribution-shape statistics and their statistical significance should be interpreted with caution when comparing merger and control populations, as the sensitivity of statistical tests generally increases with sample size.

Overall, these distribution-shape statistics support the trends identified from the \Doh\ distributions in Section \ref{sec_oh_spx} and Figure \ref{fig_dist_oh}. The early- and late-stage major-merger populations are predominantly characterized by negative skewness, quantitatively supporting the presence of extended low-\Doh\ tails associated with metallicity dilution. Although departures from Gaussianity are common across both merger and control populations, the Dip tests provide no significant evidence for multimodality. Thus, the extended low-\Doh\ tails in the major-merger populations reflect asymmetry in the metallicity-offset distributions rather than the emergence of distinct metallicity populations..

\end{CJK}
\end{document}